\documentclass[aps,prb,twocolumn,amsmath,amssymb,floatfix,superscriptaddress,longbibliography]{revtex4-2}
\usepackage{natbib}
\usepackage{graphicx,xcolor,multirow,bm,hyperref}
\usepackage{float}
\usepackage{multirow}
\usepackage[ignoreunlbld,norefs,nocites]{refcheck}
\begin{document}

\title{Quantum Monte Carlo description of correlated electrons in two-dimensional FeSe}
\author{Sam Azadi}
\email{sam.azadi@manchester.ac.uk}
\affiliation{Department of Physics and Astronomy, University of Manchester, Oxford Road, Manchester M13 9PL, United Kingdom}
\author{A.\ Principi}
\affiliation{Department of Physics and Astronomy, University of Manchester, Oxford Road, Manchester M13 9PL, United Kingdom}
\author{R. V. Belosludov}
\affiliation{Institute for Materials Research, Tohoku University, Sendai 980-08577, Japan}
\author{T. D. K\"{u}hne}
\affiliation{Center for Advanced Systems Understanding, Untermarkt 20, D-02826 G\"orlitz, Germany}
\affiliation{Helmholtz Zentrum Dresden-Rossendorf, Bautzner Landstra{\ss}e 400, D-01328 Dresden, Germany}
\affiliation{TU Dresden, Institute of Artificial Intelligence, Chair of Computational System Sciences, N\"othnitzer Stra{\ss}e 46 D-01187 Dresden, Germany}
\author{M.\ S.\ Bahramy}
\affiliation{Department of Physics and Astronomy, University of Manchester, Oxford Road, Manchester M13 9PL, United Kingdom}
\date{\today}
\begin{abstract}
An interesting question in physics is how the correlation energy of atoms evolves upon forming a solid. Here, we address this problem for a specific case of double-layer FeSe. We used many-body wavefunction-based quantum Monte Carlo (QMC) techniques to compute the correlation energies of double-layer FeSe with different geometrical configurations and compared them with those of isolated Fe and Se atoms. Variational and diffusion QMC calculations were carried out with Slater–Jastrow trial wavefunctions, employing two alternative forms for the homogeneous two-body pair-correlation term. The ground-state energy was obtained in the thermodynamic limit using two types of trial wave functions of JDFT, in which only the Jastrow factor is optimized while the Slater determinant is derived from the local density approximation, and JSD, where both the Jastrow factor and the Slater determinant are optimized simultaneously. Our results indicate that the correlation energy of double layer FeSe at the thermodynamic limit is mainly determined by the atomic contributions, with the bonding between atoms playing a comparatively minor role in it. After optimizing the interlayer separation of double-layer FeSe under tensile strain, we analyze the correlation energy as a function of strain and separation. We found that with increasing tensile stretch and interlayer spacing, the correlation energy of double-layer FeSe stochastically approaches that of its constituent atomic fragments.
\end{abstract}

\maketitle
\section{Introduction}
The electron-electron (el-el) many-body correlations in magnetic two-dimensional (2D) systems, including iron-based superconductors (FeSCs), play a fundamental role in understanding their structural, magnetic and exotic electronic properties such as unconventional superconductivity \cite{HuaWang2022,Gibertini2019,Jenkins2022,Gong2017,Kontani2010,Bohmer2018,Medvedev2009}. In two-dimensional FeSCs, electronic interactions are moderately strong, placing them in an intermediate coupling regime between weak coupling and strong correlation effects found in cuprates\cite{Kang2024}. The confinement of electrons in two-dimensional FeSCs enhances correlation effects due to the reduced screening of Coulomb interactions between electrons that introduce magnetic anisotropy, which can be tuned externally by electric fields, strain, photoexcitation, and chemical doping\cite{Dagotto2013, Dai2015,Huang2020,Kothapalli2016,Gong2017,Huang2017}. Renormalization of the electronic band structure and the increase in the effective mass of electrons, which is observed in angle-resolved photoemission spectroscopy (ARPES) and quantum oscillation experiments\cite{Zhang2022}, can be enhanced by the el-el correlation interaction. Deviations from Fermi-liquid behavior \cite{Landau} in FeSCs are introduced by electronic correlations, especially near quantum critical points where magnetic fluctuations dominate. This non-Fermi liquid behavior is reflected in unusual temperature-dependent properties $\rho \sim T^n , n >2$ and other anomalous properties \cite{Tazai2023}. Competing orders, such as charge density waves or nematic order, which often coexist or compete with superconductivity, can be stabilized due to correlation effects\cite{Chubukov2016,Yamakawa2016,Onari2016}. Nematicity, characterized by the break of rotational symmetry, is strongly influenced by many-body el-el correlations in FeSCs\cite{Fernandes2014}.  Two-dimensional FeSCs provide an ideal platform for studying the interplay between electron correlations and reduced dimensionality, bridging the gap between weakly correlated systems (such as conventional superconductors) and strongly correlated systems (such as cuprates) \cite{Dai2015,Gibertini2019,HuaWang2022,Gong2017}. In this work, many-body el-el correlations in two-dimensional iron selenide (2D-FeSe) are calculated using continuous quantum Monte Carlo methods in real space.

Two-dimensional iron selenide (2D-FeSe) provides a versatile platform for exploring quantum phenomena and exotic phases of matter within the family of iron-based superconductors \cite{Dai2015,McQueen2009,Alloul2009,MYi2013,ZYin2011,MMa2017}. 2D-FeSe exhibits unique electronic and magnetic behaviors, even within the one-particle approach \cite{Azadi2025}, suggesting a complex interplay between nematicity, magnetism, and superconductivity\cite{Yamakawa2016}. The bulk FeSe undergoes a structural phase transformation from tetragonal to orthorhombic at approximately $\sim90$ K \cite{Hsu2008,McQueen2009,Wang2020}, without any finite magnetic order under ambient conditions. Its critical superconducting transition temperature is $\sim 8$ K \cite{Hsu2008}, which can increase to $\sim37$ K under a hydrostatic pressure of $\sim6$ GPa \cite{Medvedev2009,Okabe2010}. Correlation interactions significantly renormalize electronic bands in FeSe and enhance effective masses by factors $\sim$3–5 compared to band theory predictions\cite{Tomczak2012}. Bulk FeSe undergoes a nematic phase transition at $\sim$90 K, breaking the rotational symmetry of the crystal and electronic structure, driven by orbital-dependent electronic correlations, with a strong interaction between orbitals $d_{xz}$ and $d_{yz}$\cite{McQueen2009}. However, in 2D-FeSe, the nematic phase can be suppressed, possibly due to enhanced quantum fluctuations that may contribute to the high $T_c$ by shifting the balance between competing orders\cite{XLong2020}. 

In this work, we focus on double-layer 2D-FeSe. The electronic properties of double-layer FeSe can be influenced by interlayer separation, which is governed by long-range van der Waals (vdW) interactions \cite{Ding2021}. Controlling this separation, for example, through isothermal compression, allows tuning of various emergent features, such as the coupling between nematicity and magnetism \cite{Kothapalli2016}. The coexistence of competing orders, namely strong localized interactions and weak vdW forces in the double-layer FeSe system, presents unique opportunities to explore exotic quantum phenomena, with implications for both fundamental science and technological applications.  Tensile strain applied within the Fe-Se plane can affect the electronic structure and electron correlation in double-layer FeSe. This tuning mechanism alters orbital overlap, bandwidth, and hybridization, making it a useful tool for manipulating correlated phases in double-layer FeSe. Our main tools in this study are the real space variational and diffusion quantum Monte Carlo methods, which can accurately describe the correlation-driven phenomena. 

Real-space Variational Quantum Monte Carlo (VMC) and Diffusion Quantum Monte Carlo (DMC) have proven to be highly successful in studying electronic correlation effects in transition metal compounds and capturing weak vdW interaction in 2D systems \cite{Kolorenc2010, Koloren2011, Dubecky2016, Koloren2008,Wines2025}, including systems like FeSe\cite{Busemeyer2016}, due to the fact that they directly address the many-body nature of electron interactions, which are critical in correlated systems with partially filled $d$ orbitals. Traditional mean-field approaches, such as density functional theory (DFT) with local density approximation (LDA) or generalized gradient approximation (GGA), often fail to capture strong correlations. VMC and DMC go beyond mean-field approximations by directly solving the many-body Schr\"{o}dinger equation using the stochastic techniques. Both VMC and DMC allow for systematic improvement of accuracy by refining trial wave functions and optimizing variational parameters using variance and energy minimization\cite{Umrigar2007}. 

In variational Monte Carlo (VMC) calculations, the expectation value of a many-electron Hamiltonian is calculated with respect to a trial wave function (WF) that can be of arbitrary complexity \cite{Matthew2001,BeccaSorella}. In diffusion Monte Carlo (DMC) simulations, we simulate a process controlled by the Schr\"{o}dinger equation in imaginary time to project out the ground-state component of an initial WF. We use the fixed-node approximation to impose Fermionic antisymmetry \cite{Anderson1976}. VMC provides detailed insights into the role of electronic correlations by using explicitly many-body-correlated trial wavefunctions with well-optimized Jastrow factors. DMC improves upon VMC by stochastically projecting the trial wave function onto the exact ground-state wavefunction, minimizing errors from the trial wave function. DMC is one of the most accurate methods for strongly correlated systems. 

We present the results which are obtained using both the VMC and DMC methods. VMC provides a controlled variational estimate of the ground-state energy using trial wave functions with systematically tunable accuracy. It allows us to explore how different forms and parameterizations of the wave function affect observables, making it ideal for analyzing the sensitivity of the correlation energy to nodal structure, electron localization, and bond-length variation. DMC, by projecting out the ground state (within the fixed-node approximation), provides more accurate estimates of total and correlation energies, less sensitive to the details of the trial wave function beyond the nodal surface. It serves as our primary source of benchmark-quality results. Using both methods allows us to (i) validate the robustness of physical trends across two levels of theory, (ii) quantify the impact of wave function quality on the results, and (iii) assess the affect of the nodal surface on QMC results.

\section{Method}\label{sec:method}
We provide details of our QMC simulations for double-layer FeSe in this section. The main ingredient of our QMC calculations is the many-body wavefunction with the resonance valence bond (RVB) format defined as the product of a Jastrow factor $J$ and an antisymmetrized geminal power (AGP) determinant part $\Psi_{\text{AGP}}$ \cite{Marchi2009,Azadi2025} as implemented in TurboRVB\cite{TurboRVB}. The determinant part is:
\begin{equation}
    \label{eq:AGP}
    \Psi_{\text{AGP}} (\mathbf{R}) = \mathcal{A} \Pi_{i=1}^{N_{\downarrow}} \phi(\mathbf{r}_{i}^{\uparrow}, \mathbf{r}_{i}^{\downarrow})  
\end{equation}
where $\mathcal{A}$, $\mathbf{R} = \left \{ \mathbf{r}_{1}^{\uparrow}, \cdots, \mathbf{r}_{N_{\uparrow}}^{\uparrow}, \mathbf{r}_{1}^{\downarrow},\cdots, \mathbf{r}_{N_{\downarrow}}^{\downarrow} \right \} $, and $\phi(\mathbf{r}_{i}^{\uparrow}, \mathbf{r}_{i}^{\downarrow}) = \phi(\mathbf{r}_{i}^{\downarrow}, \mathbf{r}_{i}^{\uparrow} )$, are the antisymmetrization operator, the $3N$-dimensional vector of electron coordinates, and a symmetric orbital function describing the singlet pairs, respectively. $N$, $N_\uparrow$ and $N_\downarrow$ are the total number of electrons, spin-up and spin-down electrons, respectively $N_\uparrow = N_\downarrow =N/2$. 
The pairing function in eq.~\ref{eq:AGP} is expanded in terms of molecular orbitals (MOs)
\begin{equation}
 \phi(\mathbf{r}^{\uparrow}, \mathbf{r}^{\downarrow}) = \sum_{i=1}^{N/2} \psi_i^{MO}(\mathbf{r}^{\uparrow}) \psi_i^{MO}(\mathbf{r}^{\downarrow}),
\end{equation}
The MOs are expanded in a Gaussian single-particle basis set ${\chi}$ centered on the atomic position $\psi_i^{MO}(\mathbf{r}) = \sum_j \beta_{ij} \chi_{j}(\mathbf{r})$  \cite{Marchi2009,OzoneTurboRVB}. If not mentioned, our results are obtained using an uncontracted Gaussian basis of $8s6p4d1f$ and $6s4p2d$ orbitals for Fe and Se, respectively. Initial values of the Gaussian orbitals chosen from the cc-pVTZ \cite{ccp} basis set. The core electrons of the Fe and Se atoms were replaced by correlation consistent effective core potentials (ccECPs)\cite{ccECP1,ccECP2}. Our simulation cell is subject to two-dimensional periodic boundary conditions in the $xy$ direction with standard Ewald summations and the $\Gamma$-point to calculate the Coulomb interaction. The variational parameters in our $AGP$ wave function, which is in fact the single Slater determinant, as the maximum number of MOs equals the number of electron pairs, are $\beta_{ij}$, and the exponents of the uncontracted Gaussian basis set ${\chi}$.  The initial MOs were obtained using the density functional approach \cite{Azadi2010} with the local density approximation (LDA) \cite{lda} using the same uncontracted Gaussian basis set described above. 

The Jastrow term in the trial wavefunction is responsible for the dynamic correlation between electrons and includes a homogeneous two-body factor $J_{\text{2b}}$, which is a function of the relative distance between two electrons, and a non-homogeneous three-body term $J_{\text{3b}}$ which is a function of the atomic position, defined as:
\begin{equation}
    J_{\text{2b}}= exp (\sum_{i<j} u(r_{ij})) 
\end{equation}
\begin{equation}
    J_{\text{3b}} = exp (\sum_{i<j} f(\mathbf{r}_i, \mathbf{r}_j)) ; f(\mathbf{r}_i, \mathbf{r}_j) = \sum_{ablm} g_{lm}^{ab} \chi_{al}(\mathbf{r}_i) \chi_{bm}(\mathbf{r}_j)
\end{equation}
where $g_{lm}^{ab}$ are optimizable parameters. The three-body electron-ion-electron is defined by the diagonal matrix elements $g^{aa}$. The off-diagonal $g^{ab}, a\neq b$ matrix elements define the four-body electron-ion-electron-ion terms that were not used in our study.  $f(\mathbf{r}_i, \mathbf{r}_j)$ is a two-electron coordinate function expanded using one-particle basis sets, $i, j$ are electron indices, and $r_{ij}$ are electron-electron distances. For FeSe simulations, we used two different forms for the two-body homogeneous part $u_{F} = r/(2(1+ar))$ \cite{Fahy90} and $u_{C} = \frac{b}{2}(1-exp(-r/b))$ \cite{Ceperley78}, where $r$ is the relative distance between two electrons, $a$ and $b$ are variational parameters. For Jastrow single-particle orbitals, we used uncontracted Gaussian basis sets of $3s3p1d$ and $3s2p$ for Fe and Se, respectively. The DMC energies are obtained using the time step 0.01 a.u., the 3200 walkers (configurations), and the locality approximation for the pseudopotential\cite{Casula2006}.   

\section{Results and discussions} \label{sec:results}
\subsection{Correlation energy of Fe and Se atoms}
To validate our HF approach, we calculated the correlation energy $E_{corr}=E_{QMC}-E_{HF}$ of the Fe and Se pseudo-atoms, where $E_{QMC}$ is the total energy of the atom obtained by the VMC and DMC methods, and $E_{HF}$ is the HF energy calculated by solving the Kohn-Sham Hamiltonian with \%100 exchange interaction and \%0 correlation weight factor \cite{Azadi2010}. We used large Gaussian basis sets of $10s10p7d2f1g$ and $9s9p6d1f$ for Fe ans Se atoms, respectively, to reach the complete base set (CBS) limit.  We performed QMC calculations using the same basis set as the CBS-HF and also the basis set used for the FeSe calculations ($8s6p4d1f$ and $6s4p2d$ for Fe and Se, respectively) reported in the next section. The Slater determinant was obtained using both LDA and HF methods and $u_F$ was used for the two-body homogeneous part of the Jastrow term.

Our results for the Fe atom are listed in Tab.~\ref{tab:atom_Fe}. Only the Jastrow term was optimized in the JDFT and JHF WFs. The results in Tab.~\ref{tab:atom_Fe} were obtained using uncontracted Gaussian basis sets of $3s3p1d$ for the Jastrow factor which is the same as the FeSe system. 
\begin{table}[]
    \footnotesize
    \centering
    \begin{tabular}{|ccccc|}
    \hline\hline
     BS-WF             & VMC-JDFT  & VMC-JSD  & DMC-JDFT  & DMC-JSD    \\
    \hline     
     $8s6p4d1f$-LDA    & -50.94(5) & -62.38(4) & -65.04(6) & -73.17(6)   \\
     $10s10p7d2f1g$-LDA& -67.95(1) & -68.07(1) & -72.42(2) & -72.69(2)    \\
    \hline     
     BS-WF            &  VMC-JHF & VMC-JSD & DMC-JHF & DMC-JSD   \\
     \hline
     $8s6p4d1f$-HF    & -50.02(4)& -66.14(3)& -64.92(6) & -74.72(6) \\
     $10s10p7d2f1g$-HF& -67.37(2)& -67.80(2)& -72.42(3) & -72.69(3) \\
     \hline\hline
    \end{tabular}
    \caption{The correlation energy of Fe pseudoatom (ccECP) in mHa/el obtained using $8s6p4d1f$ and $10s10p7d2f1g$ (CBS) basis sets, and LDA/HF generated Slater determinant for the WF. The correlation energy of all electron atom obtained by RPA and QC methods are -68.2 \cite{Burke2016}, and -47.5 \cite{McCarthy}, respectively. The DMC energies were calculated using time step $0.01$ a.u. }
    \label{tab:atom_Fe}
\end{table}

\begin{figure}
    \centering
    \includegraphics[width=0.75\linewidth]{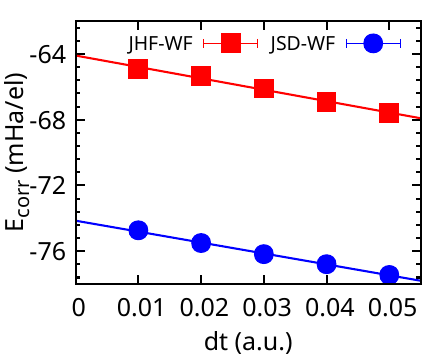}
    \caption{DMC correlation energy of Fe as a function of time step. Energies were calculated using $8s6p4d1f$-HF WF and ccECP.}
    \label{fig:dmcvsdt}
\end{figure}

Our QMC results for the correlation energy of the Fe pseudoatom agree relatively well with the correlation energy of the all-electron Fe atom obtained by the Random Phase Approximation (RPA) and Quantum Chemistry (QC) methods: -68.2 \cite{Burke2016} and -47.5 \cite{McCarthy}, respectively. The DMC-JSD-$E_{corr}$ results show a negligible dependence on the size of the basis set and whether LDA or HF is used to generate the Slater determinant.  We investigated the DMC time-step error using $8s6p4d1f$-HF WF. Fig.\ref{fig:dmcvsdt} shows that the DMC time-step error in the $E_{corr}$ results reported in Tab.~\ref{tab:atom_Fe} is negligible. The DMC correlation energies at the zero time step obtained by the JHF and JSD WFs are -64.0(1) and -74.14(6) mHa/el, respectively. We used the same DMC time step of $0.01$  a.u. for the following FeSe calculations.

The other factor which may affect the QMC correlation energy is PP. To investigate this factor, we calculated the VMC and DMC correlation energy of the Fe atom using $8s6p4d1f$-HF WF and different PPs including ccECP\cite{ccECP1,ccECP2}, CEPP\cite{CEPP}, eCEPP\cite{eCEPP}, and ECP-Burkatzki\cite{BurkPP1,BurkPP2}. The results, which are listed in Tab.~\ref{tab:PP}, indicate negligible differences between the QMC correlation energies obtained using the studied PPs. 

\begin{table}[]
    \centering
    \begin{tabular}{|ccccc|}
    \hline\hline
    PP & VMC-JHF & VMC-JSD & DMC-JHF & DMC-JSD   \\
    \hline
    ccECP & -50.02(4)& -66.14(3)& -64.92(6)& -74.72(6) \\ 
    CEPP  & -52.13(3)& -65.65(2)& -64.60(6)& -72.71(4)\\
    eCEPP & -51.33(3)& -65.15(2)& -64.08(6)& -72.56(4)\\
    ECP   & -54.98(3)& -66.32(2)& -67.03(6)& -74.19(4)\\
    \hline\hline
    \end{tabular}
    \caption{Correlation energy of Fe pseudoatom in mHa/el obtained using VMC and DMC with different PPs and $8s6p4d1f$-HF WF.}
    \label{tab:PP}
\end{table}

The VMC and DMC correlation energies of the Se atom using ccECP and LDA-generated WF with two different basis sets are listed in Tab.~\ref{tab:atom_Se}. The correlation energies of the Se all-electron atom obtained by the RPA and QC methods are -71.5 \cite{Burke2016} and -51.2 \cite{McCarthy}, respectively. The VMC/DMC-JSD results are close to the QC all electron results.
\begin{table}[]
    \footnotesize
    \centering
    \begin{tabular}{|ccccc|}
    \hline\hline
     BS-WF           & VMC-JDFT  & VMC-JSD  & DMC-JDFT  & DMC-JSD    \\
    \hline     
     $6s4p2d$-LDA    & -14.34(3) & -44.78(1) & -15.07(6) & -44.76(3)   \\
     $9s9p6d1f$-LDA  & -44.75(2) & -45.05(1) & -44.78(3) & -45.04(2)    \\
     \hline\hline
    \end{tabular}
    \caption{The correlation energy of Se pseudoatom (ccECP) in mHa/el. The correlation energy of all electron atom obtained by RPA and QC methods are -71.5 \cite{Burke2016}, and -51.2 \cite{McCarthy}, respectively. The DMC energies were calculated using time step $0.01$ a.u. }
    \label{tab:atom_Se}
\end{table}

As we mentioned above, all the results $E_{corr}$ for the Fe and Se atoms were calculated using a large basis set to obtain the HF energy at CBS limit. In principle, a larger basis set can be used, which would affect $E_{corr}$. Due to technical difficulties, we only used Gaussian exponents smaller than 100 for CBS-HF calculations. The rest of this paper uses ccECP and the same CBS for HF calculations. The total VMC, DMC, and HF energies of Fe and Se atoms obtained using JDFT and JSD wavefunctions with different PPs are reported in the supplementary material\cite{Suppl}.

\subsection{Correlation energy of Double-layer FeSe at the thermodynamic limit}
In this section, the correlation energy of double-layer FeSe is evaluated in the thermodynamic limit by computing the ground-state total energies of three finite systems using QMC and HF, followed by extrapolation to the infinite-size limit. 

\begin{figure*}
    \centering
    \begin{tabular}{ccc}
    \includegraphics[width=0.3\linewidth]{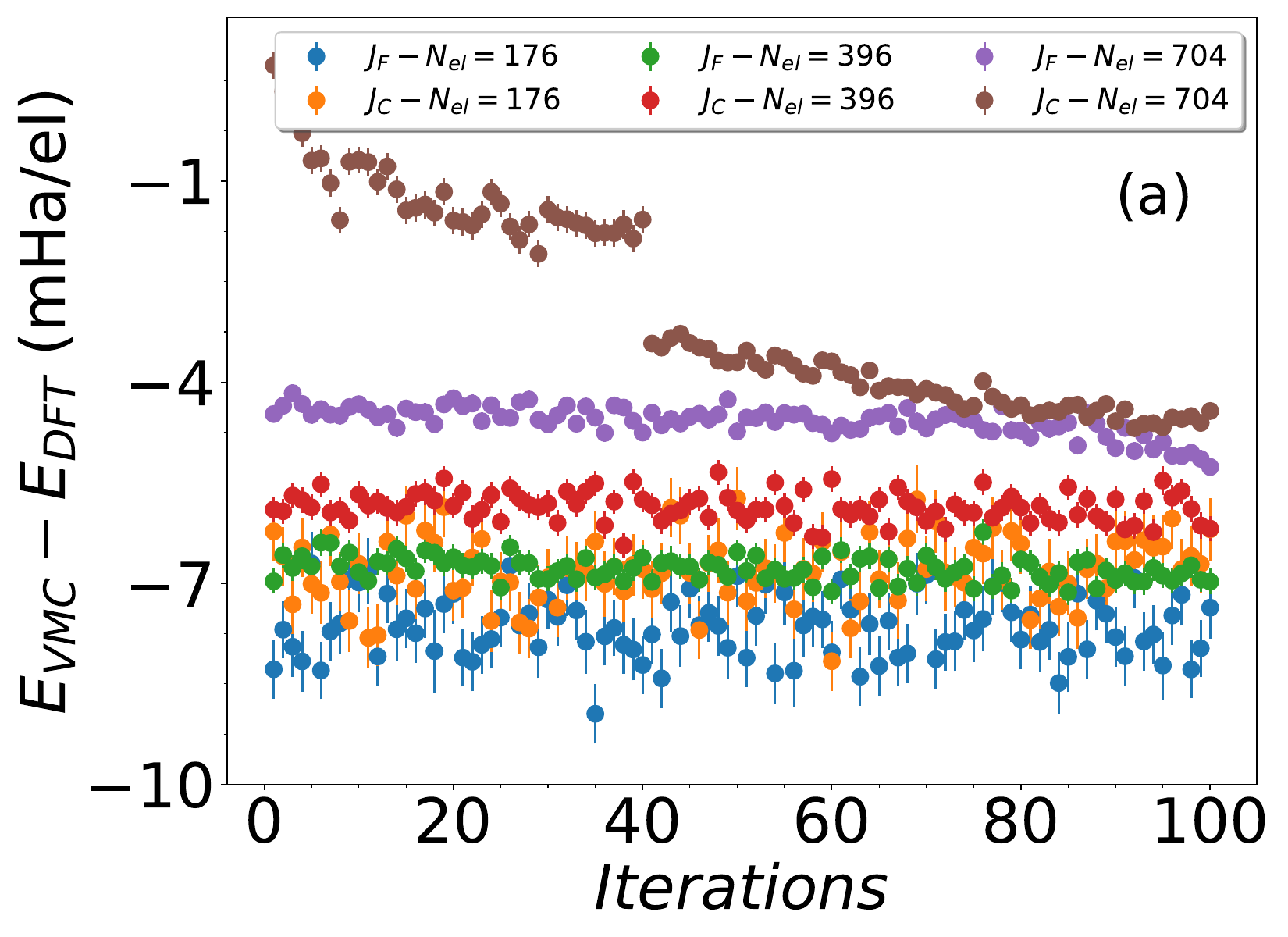} & 
    \includegraphics[width=0.3\linewidth]{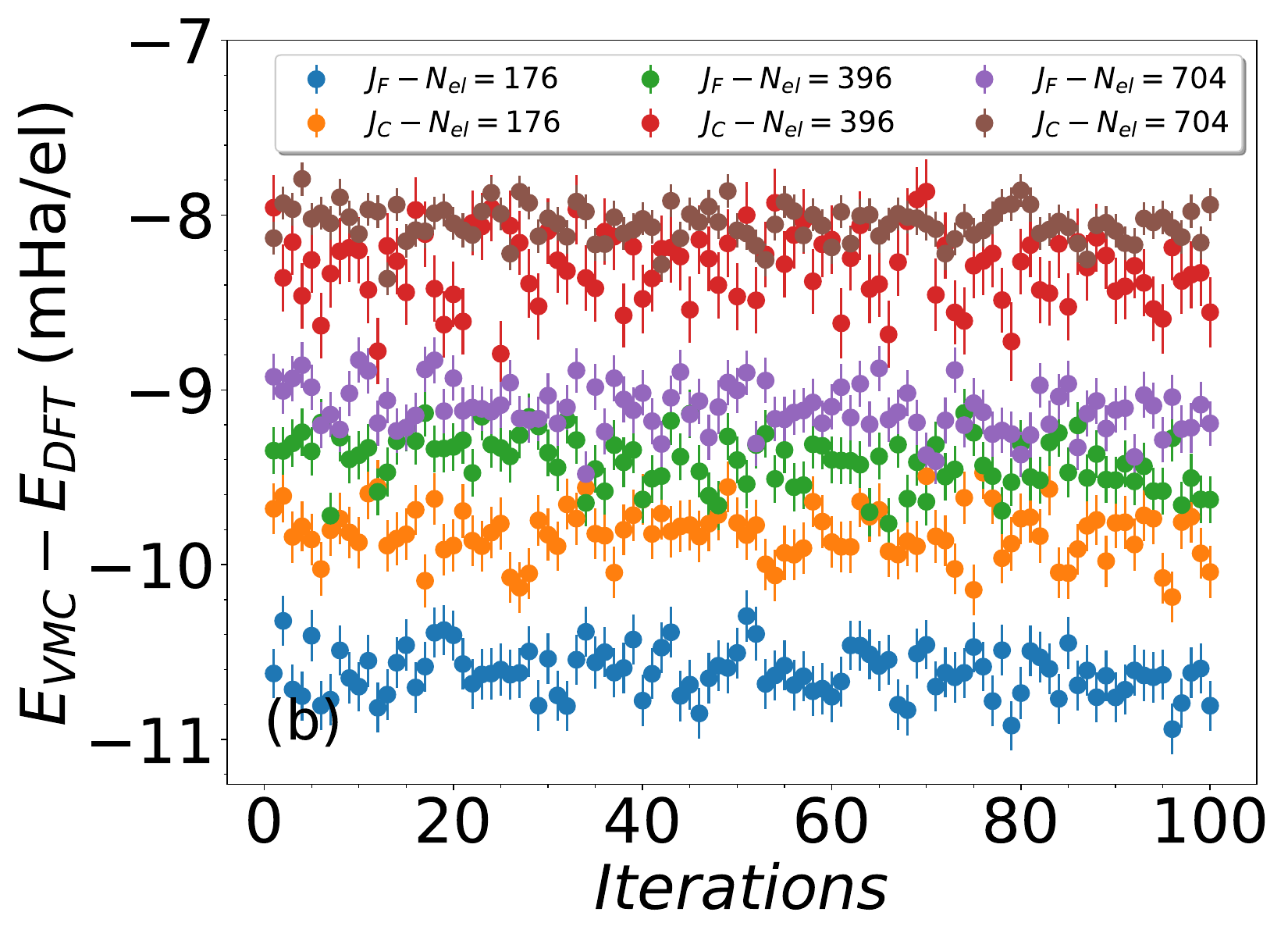} & 
    \includegraphics[width=0.3\linewidth]{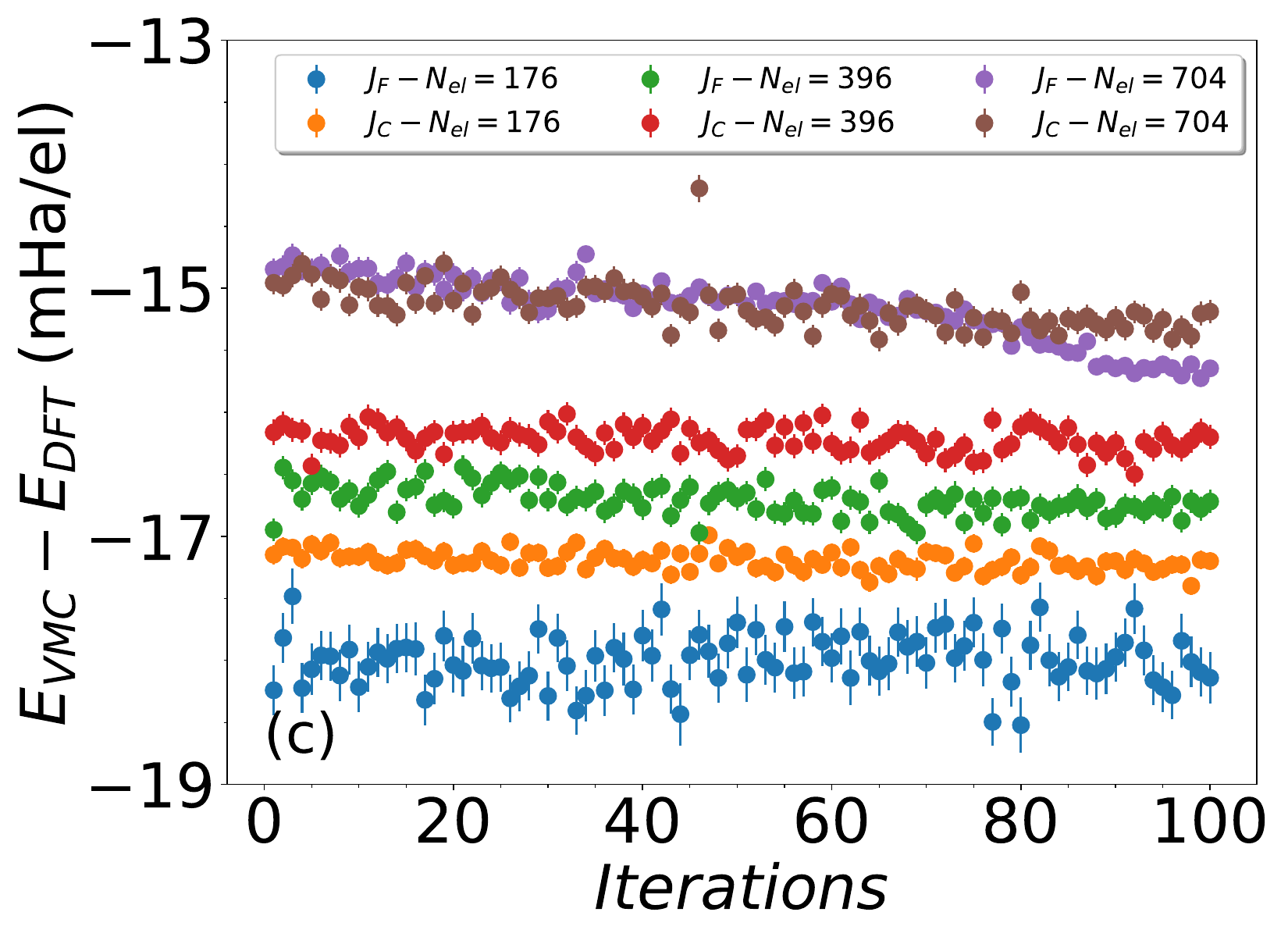} \\
    \end{tabular}
    \caption{The difference between VMC and DFT energies as a function of the number of optimization interation. (a) Only the coefficients of the Jastrow terms are optimized, (b) The exponent and Coefficients of the Jastrow term are optimized simultaneously, (c) The Jastrow and Slater determinat are optimised. Two Jastrow forms of $u_F$ and $u_C$ were used. Only the last one hundred optimization steps are shown. Three system sizes with the number of electrons in the simulation cell $N_{el}=176, 396, 704$ were considered.}
    \label{fig:opt}
\end{figure*}

We focused on optimizing the trial wave function for each system size and lattice parameter before performing the final VMC and DMC simulations. The trial wave function is crucial in QMC simulations because it directly affects the accuracy, efficiency, and stability of the simulations, as it serves as a guiding function for importance sampling in DMC, which modifies the diffusion equation to improve convergence. A poorly optimized trial wave function can lead to large statistical fluctuations, making it difficult to obtain precise results. Optimizing the trial wave function reduces variance, leading to faster and more stable simulations. We systematically optimized the trial wave function by following three steps: (I) The Jastrow coefficients and the two-body term $u$ were optimized, as shown in Fig.~\ref{fig:opt}(a). In this step, depending on the size of the system, $\sim$ 4-8 mHa/el VMC energy obtained with respect to DFT. (II) The Jastrow function was fully optimized (Fig.~\ref{fig:opt}(b)) and the wave function was called JDFT. The energy difference between VMC and DFT is $\sim$ 8-11 mHa/el, depending on the system size. (III) The Jastrow terms and the Slater determinant were optimized simultaneously, and the wave function was named JSD. Since the FS errors normally increase by the number of electrons in the simulation cell, the VMC energy gain with respect to DFT is smaller for larger systems. We found that following these three steps in order, meaning that the well-optimized wave function at the end of each step is used as the initial wave function for the next step, is required to reach an accuracy of mHa/el and avoid local minima. 

\begin{figure}
    \centering
        \begin{tabular}{cc}
           \includegraphics[width=0.5\linewidth]{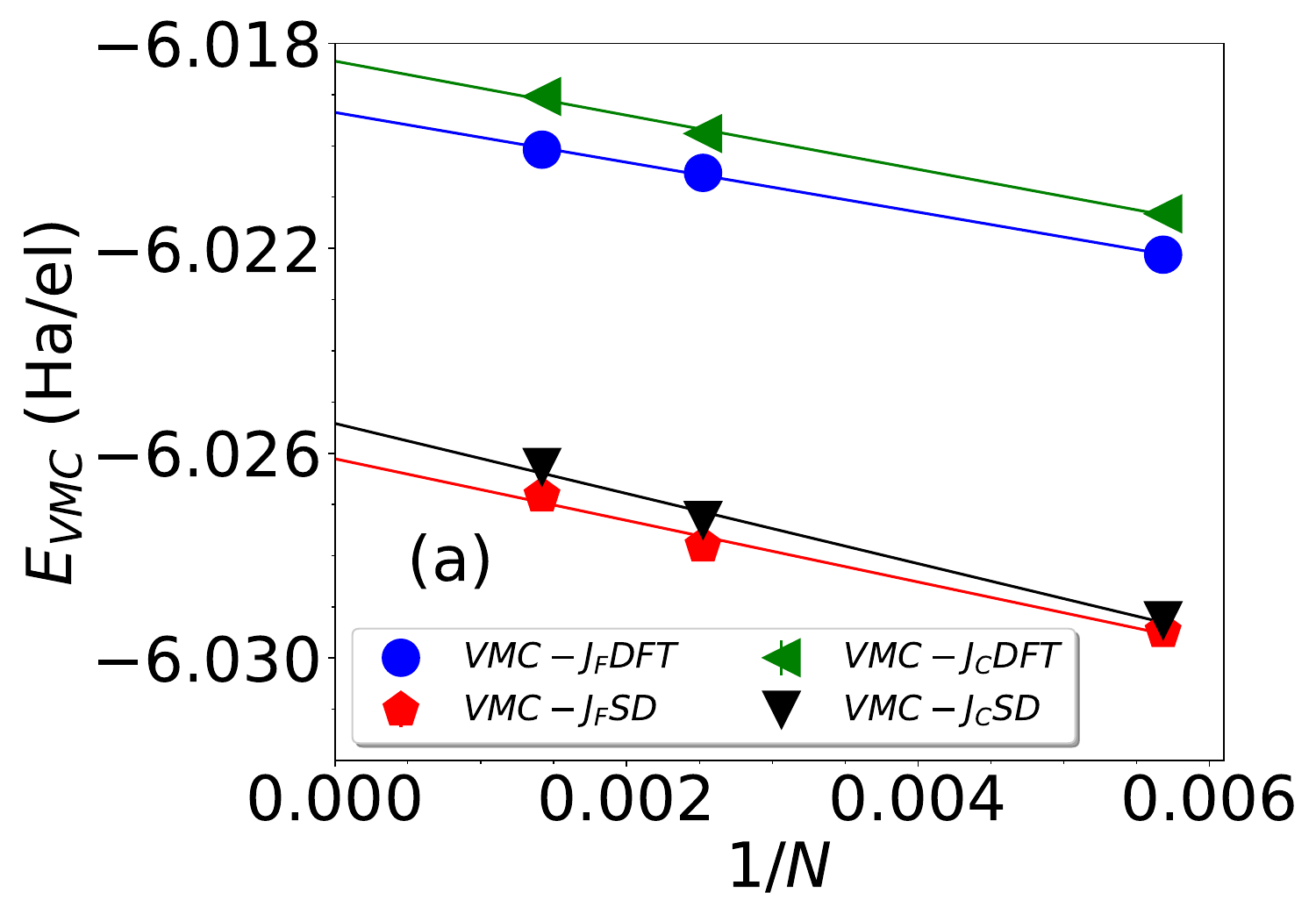}  &
           \includegraphics[width=0.5\linewidth]{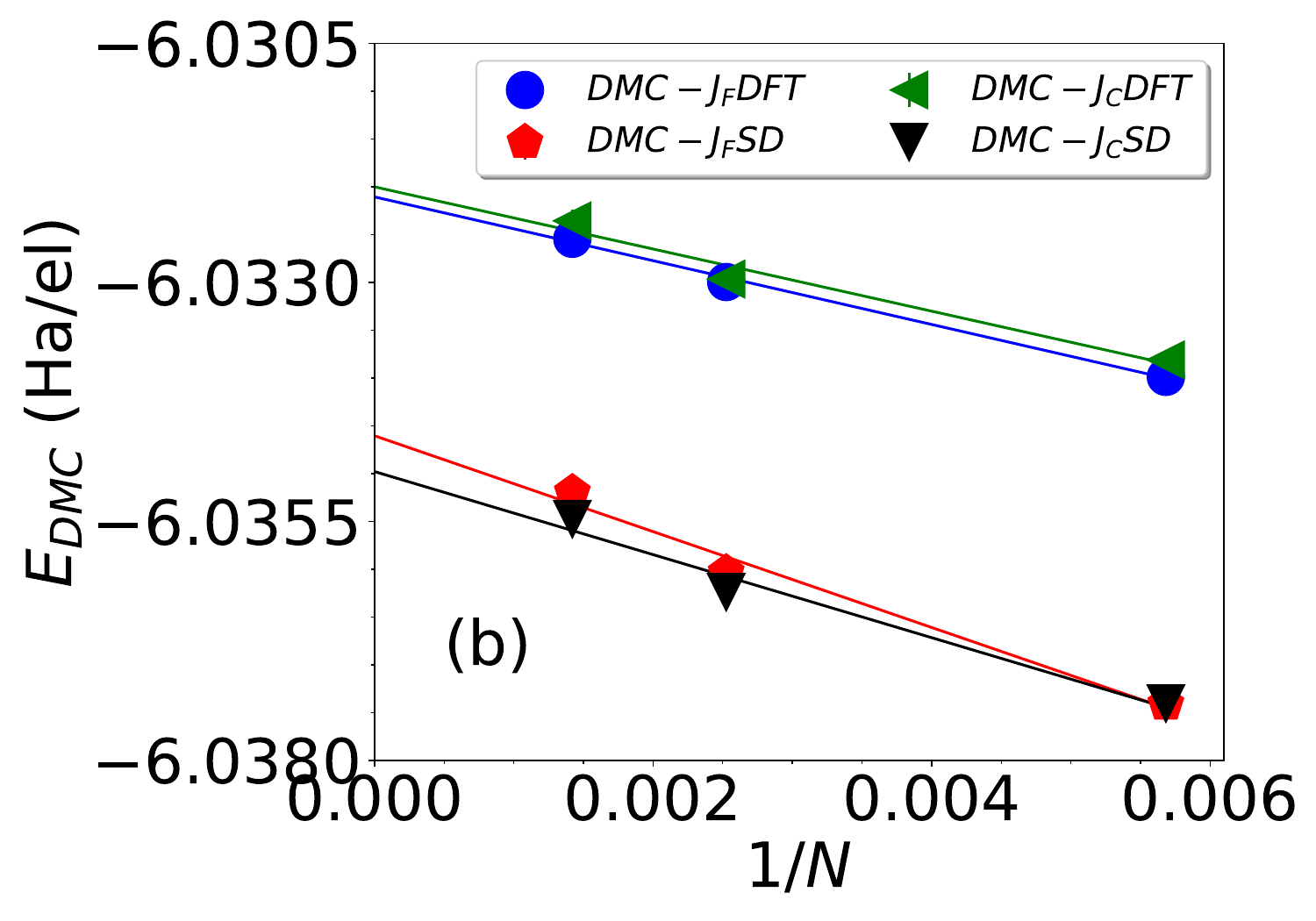}  \\
        \end{tabular}
    \caption{(a) VMC and (b) DMC energy of double layer FeSe as a function of $1/N$, where $N$ is the number of electrons in the simulation cell, obtained using $J_F$ and $J_C$ two-body el-el Jastrow factors.}
    \label{fig:FS}
\end{figure}

\begin{figure}
    \centering
    \begin{tabular}{cc}
    \includegraphics[width=0.45\linewidth]{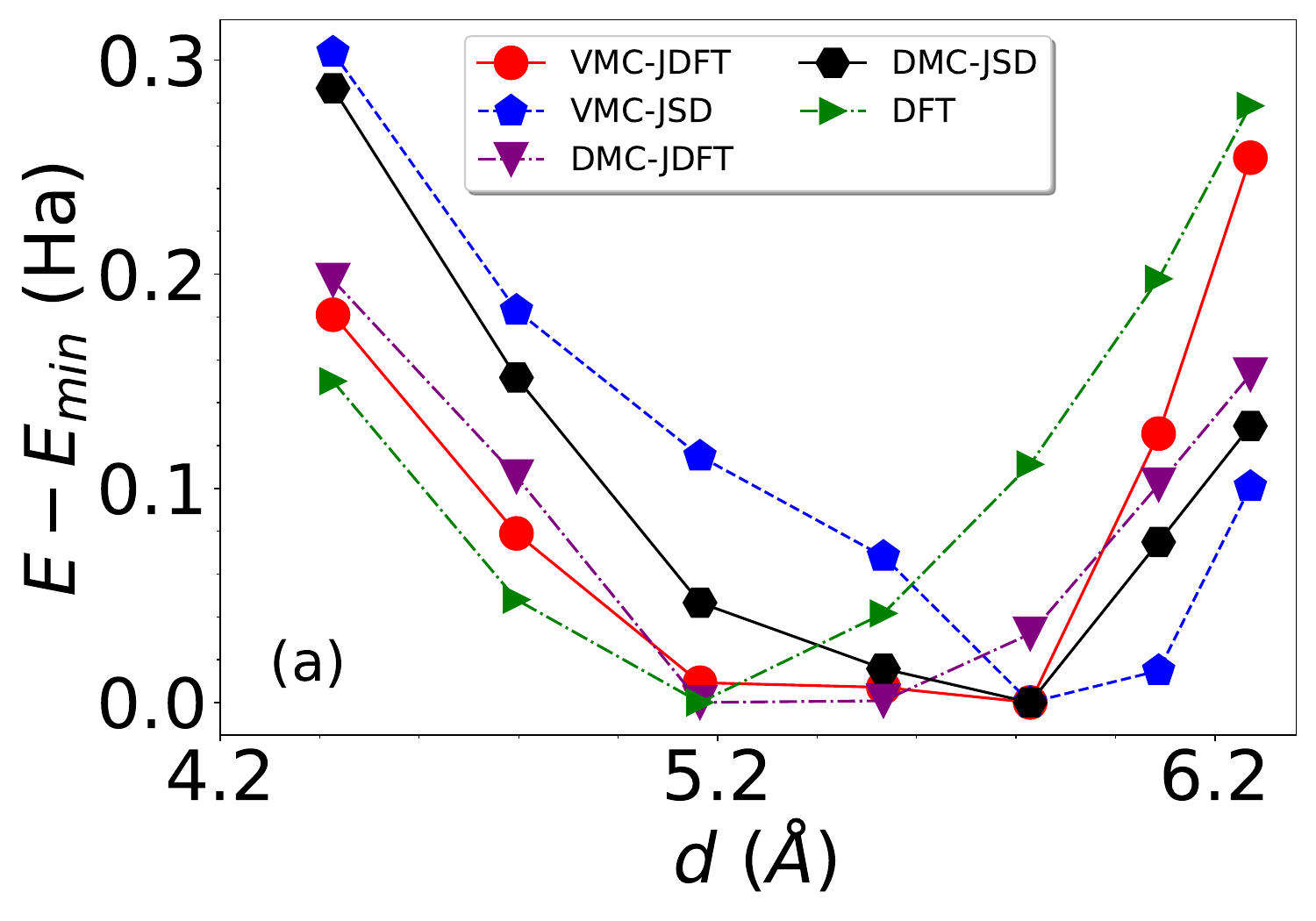} &
    \includegraphics[width=0.45\linewidth]{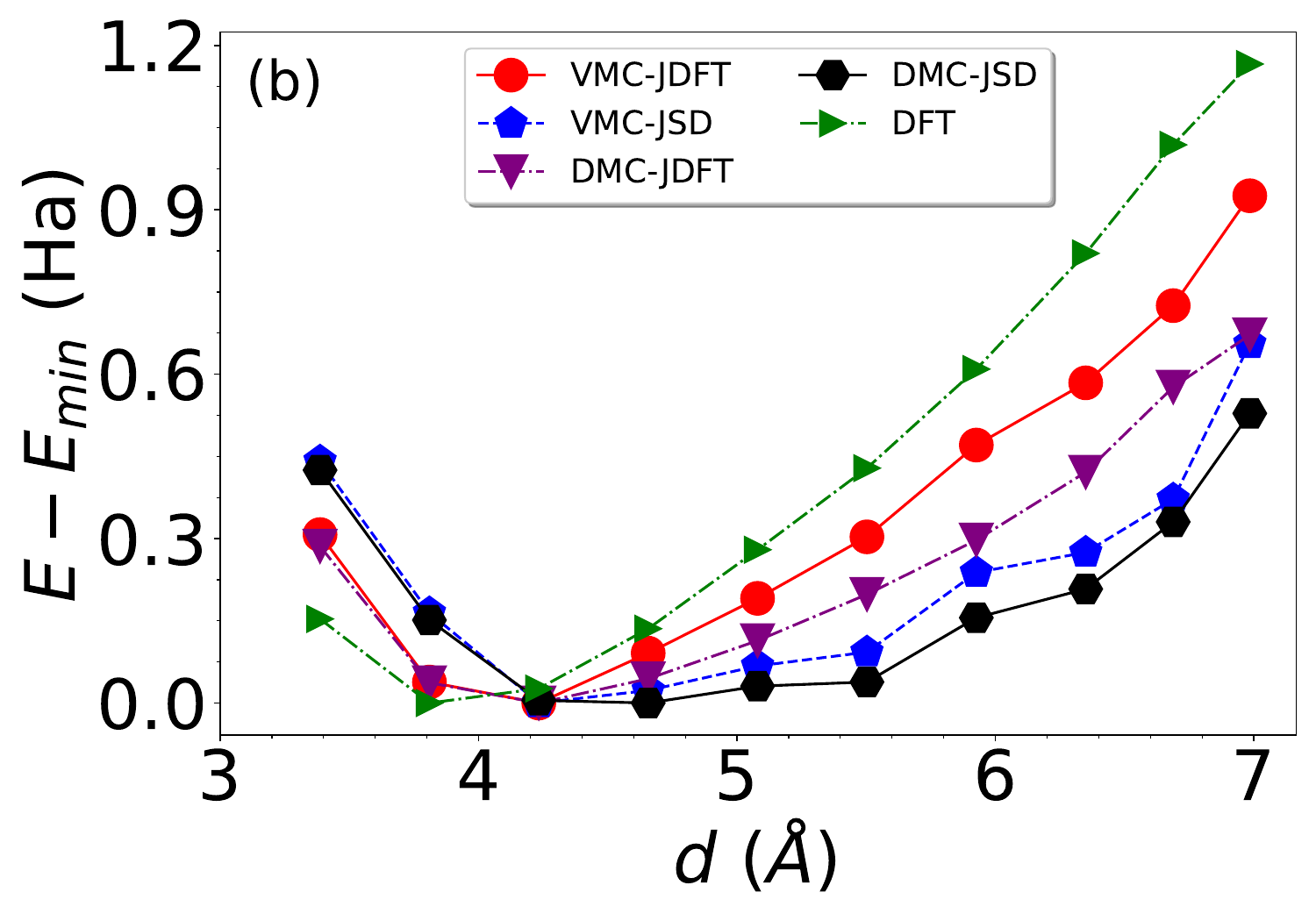} \\
    \includegraphics[width=0.47\linewidth]{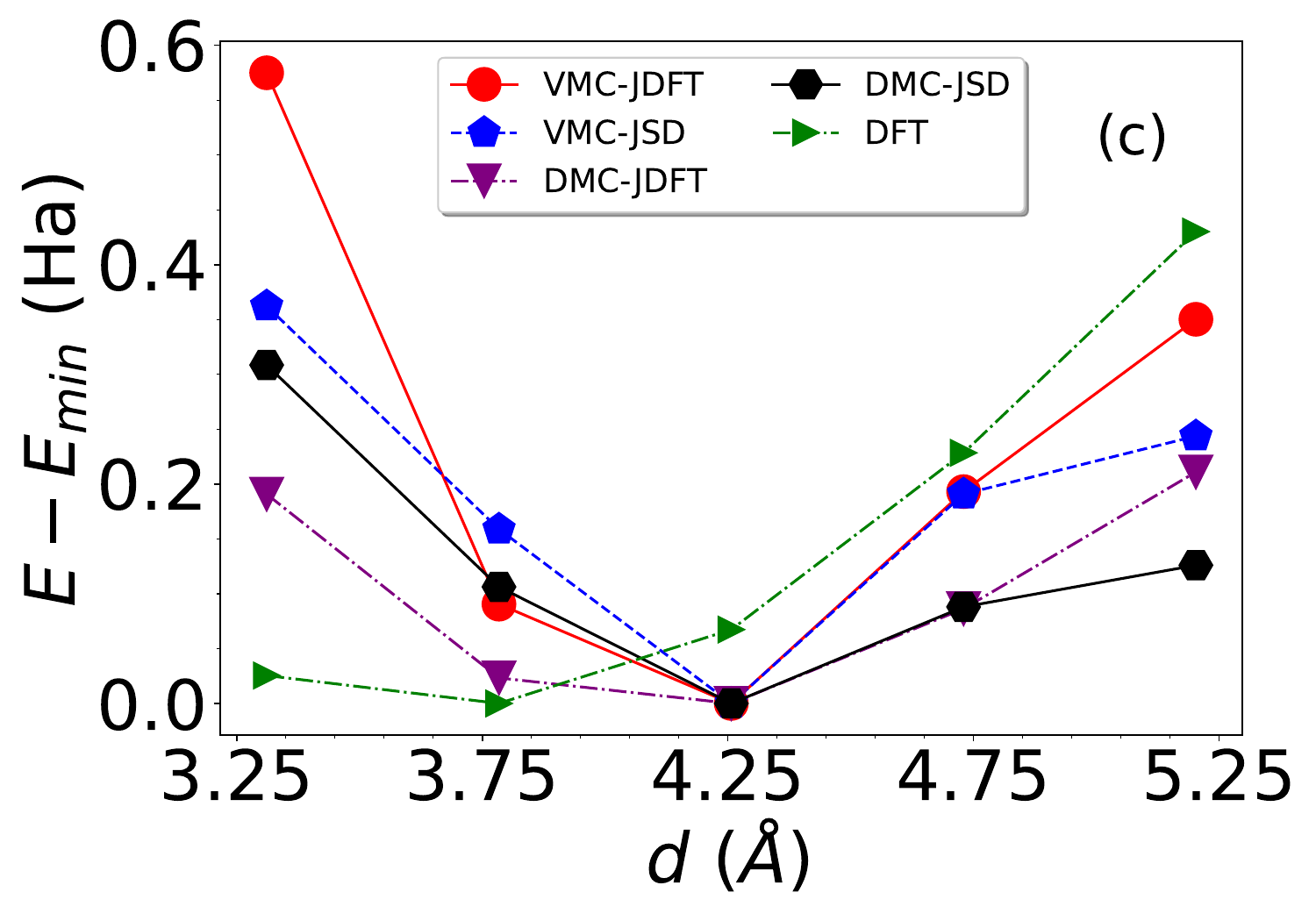} \\
    \end{tabular}
    \caption{Energy of double-layer FeSe with respect to the minimum energy $E_{min}$ as a function of separation between layers $d$. The energies are calculated using VMC and DMC with JDFT and JSD wave functions and also DFT (LDA). The energy curves are obtained for double-layer FeSe with Fe-Fe bond length (a) 3.69, (b) 4.23, and (c) 4.73~\AA.}
    \label{fig:EOS}
\end{figure}

Figure~\ref{fig:opt} shows the difference between the VMC and DFT energies during the last 100 optimization steps out of several hundreds of optimization iterations. The number of optimization steps was increased by system size. The results of the WF optimization procedure are shown for three system sizes and two-body Jastrow forms of $u_F$ and $u_C$, which are defined in Section~\ref{sec:method}.  The VMC and DMC energies obtained by the wave functions $JDFT$ and $JSD$ as a function of the system size are shown in Fig.~\ref{fig:FS}. All the results of Figs.~\ref{fig:opt} and \ref{fig:FS} were obtained for double layer FeSe with in-plane iron-iron bond length 3.69~\AA and interlayer separation 5.83 \AA~, which is the optimized value predicted by our DMC results (Fig.~\ref{fig:EOS}). The ground state energy of the system at the infinite system size limit is obtained by using linear extrapolation of the VMC and DMC energies as a function of $1/N$, where $N$ is the number of electrons in the simulation cell. The values of the VMC and DMC energies at each system size and the thermodynamic limit are presented in Table~\ref{tab:EvsN}. 

\begin{table*}
    \footnotesize
    \begin{tabular}{|c|cccc|cccc|c|}
    \hline\hline
    $N$ & VMC-$J_FDFT$& DMC-$J_FDFT$& VMC-$J_FSD$& DMC-$J_FSD$ & VMC-$J_CDFT$ & DMC-$J_CDFT$ & VMC-$J_CSD$ & DMC-$J_CSD$ & HF-CBS \\
    \hline
     $176$ &$-6.022123(8)$&$-6.03398(3)$&$-6.029479(6)$&$-6.03741(3)$&$-6.021325(8)$&$-6.033809(4)$&$-6.029264(6)$&$-6.03740(2)$ & $-5.970143$\\
     $396$ &$-6.020526(9)$&$-6.03299(5)$&$-6.027806(7)$&$-6.03603(4)$&$-6.019759(9)$&$-6.032964(6)$&$-6.027308(7)$&$-6.03624(4)$ & $-5.971588$\\
     $704$ &$-6.02007(1)$ &$-6.03254(8)$&$-6.026830(9)$&$-6.03519(8)$&$-6.01904(2)$ &$-6.03235(8)$ &$-6.02627(1)$ &$-6.03547(9)$ & $-5.973026$ \\
   $\infty$&$-6.01934(1)$ &$-6.0321(1)$ &$-6.02611(1)$ & $-6.0346(2)$&$-6.01834(3)$ &$-6.0320(2)$  &$-6.0232(1)$  &$-6.0349(3)$  & $-5.9736$\\
     \hline\hline
    \end{tabular}
    \caption{VMC and DMC energies at different system size and the thermodynamic limit obtained using $J_F$ and $J_C$ two-body el-el Jastrow term. The energy at the infinite system size limit is obtained using the linear extrapolation (Fig.~\ref{fig:FS}). The last column lists HF energy with CBS.}
    \label{tab:EvsN}
\end{table*}

Performing QMC simulations in a finite supercell with periodic boundary conditions (PBC) suffers from finite-size (FS) errors \cite{Azadi2015}. The main source of FS errors includes (i) the single-particle FS effect, which is because of the discrete $k$-point sampling in a finite simulation cell, the electron momentum states are quantized leading to errors in kinetic and exchange energies. (ii) Coulomb FS error, which is due to the issue that the long-range Coulomb interaction is artificially truncated by PBCs and therefore self-interaction and electrostatic image effects distort the interaction energy. (iii) Many-body correlation FS error mainly driven by the absence of long-wavelength collective excitations in small simulation cells. To correct FS errors, we used the Ewald summation with $\Gamma$-point and FS scaling extrapolation as shown in Fig.~\ref{fig:FS}. The single-particle FS errors in the kinetic energy part of Hamiltonian can be corrected by twist averaging \cite{Azadi2015}. The almost perfect linear behavior of the VMC and DMC energies as a function of $1/N$ suggests that the single-particle FS errors are relatively small in our simulations. 

Our VMC and DMC energies for double-layer FeSe at equilibrium configuration are obtained using two forms for two-body Jastrow functions. The pair correlation function $u_F(r)$ is more suitable in atomic chemical bond separation, while $u_C(r)$, which is large and positive for $r=0$ and goes to zero as $r\rightarrow\infty$, is useful when the atoms are at a distance greater than the variational parameter $b$. The difference between VMC energies at the thermodynamic limit obtained by the wave function $JSD$ and $u_F(r)$ and $u_C(r)$ is small. We propose the idea that using two different forms for the two-body Jastrow term and finding a good agreement between the VMC energies can be used as a test for the accuracy of wave function optimization. The DMC energies are almost independent of the form of Jastrow functions, as our results show (Fig.~\ref{fig:FS}), since the nodal surface is not directly affected by the Jastrow function. 

To investigate the effect of basis set size on our QMC energies, we increased the number of Fe-$d$ orbitals in the single-particle basis set of the Slater determinant and the Jastrow factor. The VMC and DMC energies of the double layer FeSe,  which are calculated using the determinant and Jastrow base set of $8s6pnd1f, (n=4,6)$ and $3s3pnd, (n=1,3)$, respectively, for the iron atom, are listed in table~\ref{tab:basis}. The energies are obtained for double-layer FeSe with Fe-Fe bond length of 3.69 \AA~ and interlayer separation of 5.83 \AA.  Increasing the number of Fe-$d$ orbitals in the Slater determinant basis set improves the ground-state energy more than in the Jastrow basis set. Compared with DMC, the VMC energies are more affected by the number of Fe-$d$ orbitals in the base set. More importantly, full optimization of the wave function reduces the effect of number of Fe-$d$ orbitals in the basis set as shown by the DMC-JSD energies (tab.~\ref{tab:basis}). Comparison of the DMC-JSD energies of the smallest and largest basis sets shows that reaching the mHa/el accuracy is possible by full optimization of the wave function and without increasing the size of the basis set. The DMC-JSD energy difference between the smallest and the largest basis sets is less than 0.4 mH/el. Increasing the number of $d$ orbitals may improve the flexibility of the trial wave function and allows it to better represent subtle features in orbital hybridization and correlation, as this expansion led to better improvements in VMC energy, as the difference in VMC-JSD energy between the smallest and the largest basis sets is $\sim 2.8$ mHa/el.

\begin{table}
     \footnotesize
     \centering
    \begin{tabular}{|c|cc|cc|}
    \hline\hline
    Basis   & VMC-JDFT & VMC-JSD & DMC-JDFT & DMC-JSD \\
    \hline
    $6d$-$J3d$&$-6.028753(6)$&$-6.032268(5)$&$-6.03596(1)$ &$-6.03775(1)$ \\
    $6d$-$J1d$&$-6.027060(6)$&$-6.030781(5)$&$-6.035886(9)$&$-6.037612(8)$  \\
    $4d$-$J3d$&$-6.022436(6)$&$-6.030045(5)$&$-6.034025(9)$&$-6.037344(8)$  \\     
    $4d$-$J1d$&$-6.022123(6)$&$-6.029479(5)$&$-6.033987(9)$&$-6.037408(8)$  \\     
    \hline\hline
    \end{tabular}
    \caption{VMC and DMC energies of double layer FeSe obtained using determinant single-particle orbital basis set $8s6pnd1f; (n=4,6)$ and $6s4p2d$ for Fe and Se, respectively, and Jastrow orbital $3s3pnd, (n=1,3)$ and $3s2p$ for Fe and Se, respectively. On the number of $d$ orbital is listed in the first column. Energies are in Ha/el and are calculated using two-body Jastrow $u_F$. The energies are obtained for double-layer FeSe with Fe-Fe bond length 3.69 \AA~and the distance between layers 5.83 \AA. The number of electrons in the simulation cell is 176.}
    \label{tab:basis}
\end{table}

Comparison of the difference between JDFT and JSD energies and DFT indicates that substantial energy gain can be obtained by fully optimizing the trial WF.  The results indicate that the quality of the VMC and DMC energies strongly depends on the trial wave function. The fixed node (FN) approximation, used in the DMC calculations, restricts the many-body WF to having the same nodal surface as the trial WF initially obtained by DFT-LDA. This prevents walkers from crossing into regions of opposite sign, avoiding the sign problem but introducing bias. The calculated DMC energy is an upper bound to the true ground-state energy, and the accuracy of the results depends on the quality of the WF. The effect of FN approximation on DMC energies of iron-based compounds can be large because of the complexity of the nodal surface. As our results show, the FN-DMC energy obtained using JDFT is only as good as the trial wave function obtained by LDA-DFT, which inadequately captures strong electron correlations.

We compared the correlation energy of double-layer FeSe at the thermodynamic limit with that of isolated Fe and Se atoms, as summarized in Table~\ref{tab:cooratinf}. The correlation energies were computed using ccECP and the $u_F$ form for the two-body Jastrow factor. Both VMC and DMC results obtained with the JSD wave function indicate that the average of the atomic correlation energies of Fe and Se provides a reasonable approximation for the correlation energy of the extended FeSe system. This suggests that the total correlation energy of FeSe is primarily determined by the atomic contributions, with the bonding between atoms playing a comparatively minor role in $E_{corr}$. The results in Table~\ref{tab:cooratinf} show that the magnitude of $E_{corr}$/el of FeSe lies between those of the isolated Fe and Se atoms, being smaller than in Fe but larger than in Se. This trend reflects the nature of the valence orbitals. In fact, Fe $d$ electrons are spatially localized, leading to strong onsite Coulomb interactions, while the $s$- and $p$-valence electrons of Se are more delocalized. A comparison of DMC-JDFT and DMC-JSD calculations for FeSe and Fe further reveals that the difference between DMC-JSD(FeSe) and DMC-JSD(Fe) is more than twice that between DMC-JDFT(FeSe) and DMC-JDFT(Fe). An analogous trend is observed in the VMC results obtained with the JDFT and JSD WFs. 

\begin{table}
    \centering
    \begin{tabular}{|c|cccc|}
    \hline\hline
    system & VMC-JDFT & VMC-JSD & DMC-JDFT & DMC-JSD \\
    \hline
      FeSe & -45.74(1) & -52.51(1) & -58.50(1) & -61.0(2)\\
      Fe   & -50.02(4) & -66.14(3) & -64.92(6) & -74.72(6) \\ 
      Se   & -14.34(3) & -44.78(1) & -15.07(6) & -44.76(3) \\
      \hline\hline
    \end{tabular}
    \caption{Correlation energy of bilayer FeSe at the infinite system size limit, Fe and Se atoms in mHa/el obtained using VMC and DMC methods with JDFT and JSD WFs.}
    \label{tab:cooratinf}
\end{table}

\subsection{Effects of tensile strain and interlayer separation on correlation energy}
In this section, the effects of tensile stress on the geometrical and electronic structure properties of double-layer FeSe are studied.  We also calculated the VMC and DMC energy curves of double-layer FeSe as functions of interlayer separation. We considered the experimental inplane iron-iron bond length of 3.69~\AA~and the stretched iron-iron bond lengths of 4.23, and 4.73~\AA. The energy curves, which are obtained using the JDFT and JSD wave functions, are illustrated in Fig.~\ref{fig:EOS}. All energy data points in Fig.~\ref{fig:EOS} are obtained using the basis set $8s6p4d1f/3s3p1d$ for the iron atom and N=176 electrons in the simulation cell. Comparison of the energy curves as a function of the distance between FeSe layers (Fig.~\ref{fig:EOS}) shows that the optimized separation between FeSe layers decreases as the iron-iron bond length stretches. This suggests that increasing the electron density localization as a result of in-plane stretch enlarges the attractive van der Wall interaction between two layers of double-layer FeSe.

\begin{figure}
    \centering
    \begin{tabular}{cc}
    \includegraphics[width=0.45\linewidth]{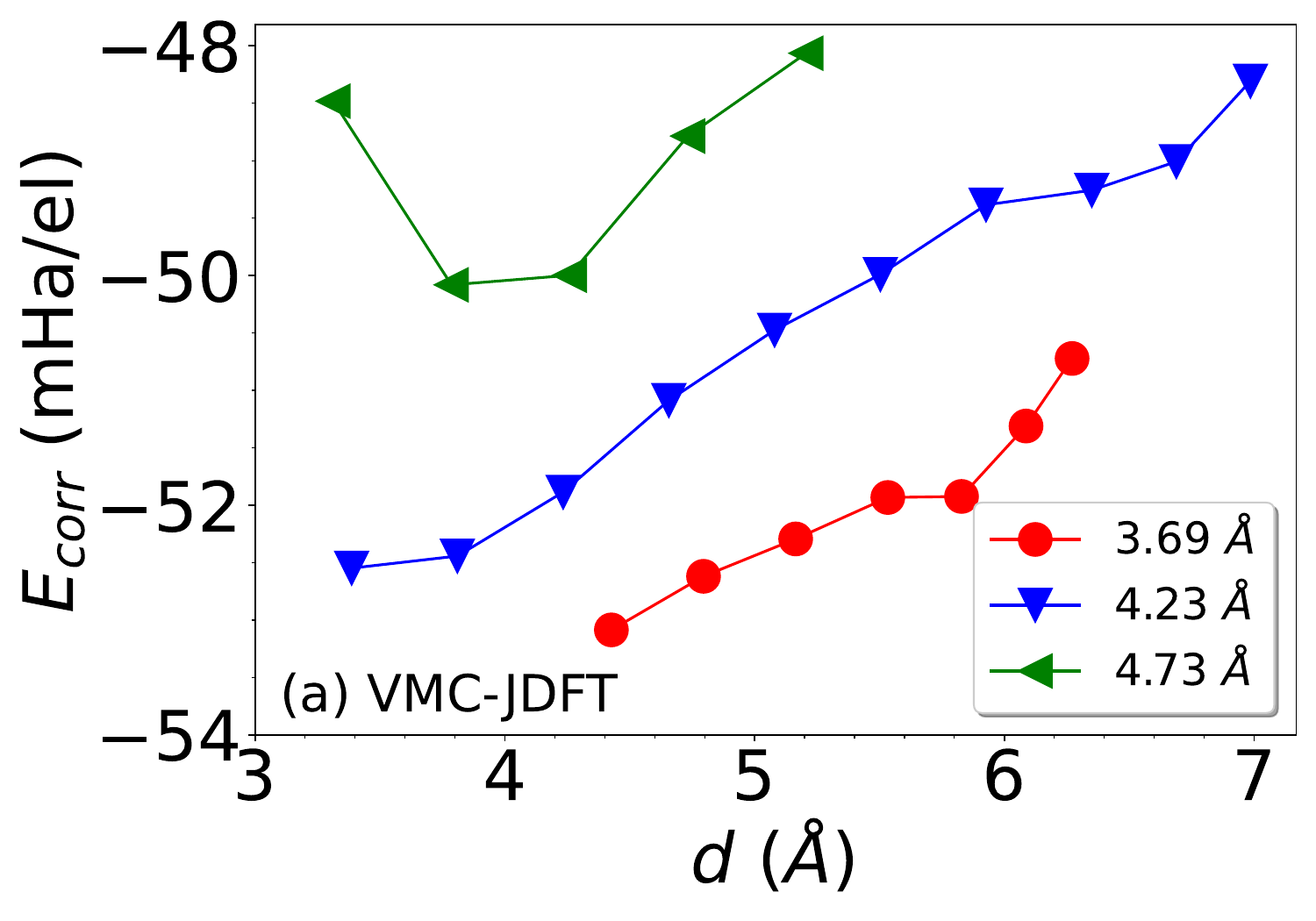}& 
    \includegraphics[width=0.45\linewidth]{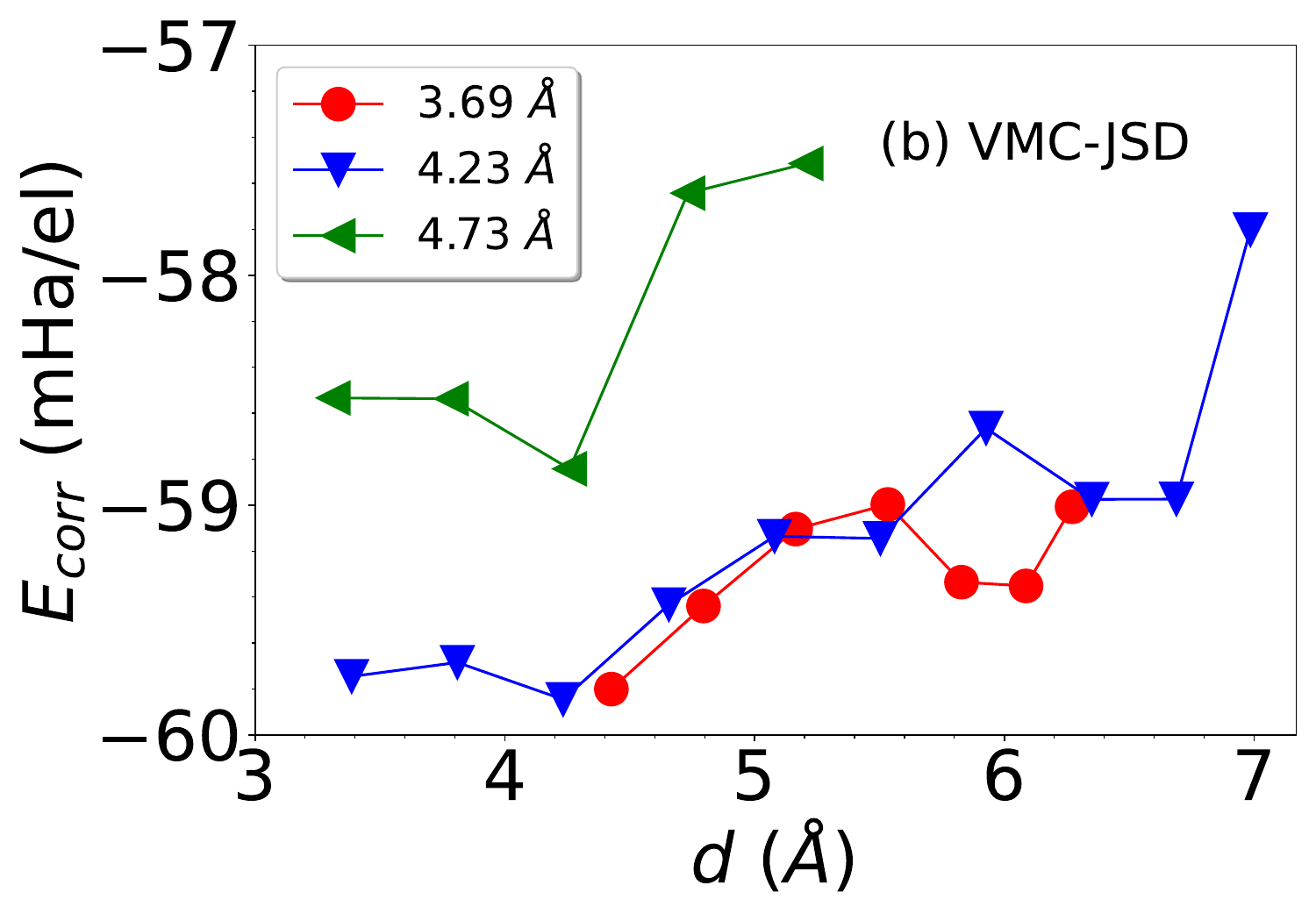}\\
    \includegraphics[width=0.45\linewidth]{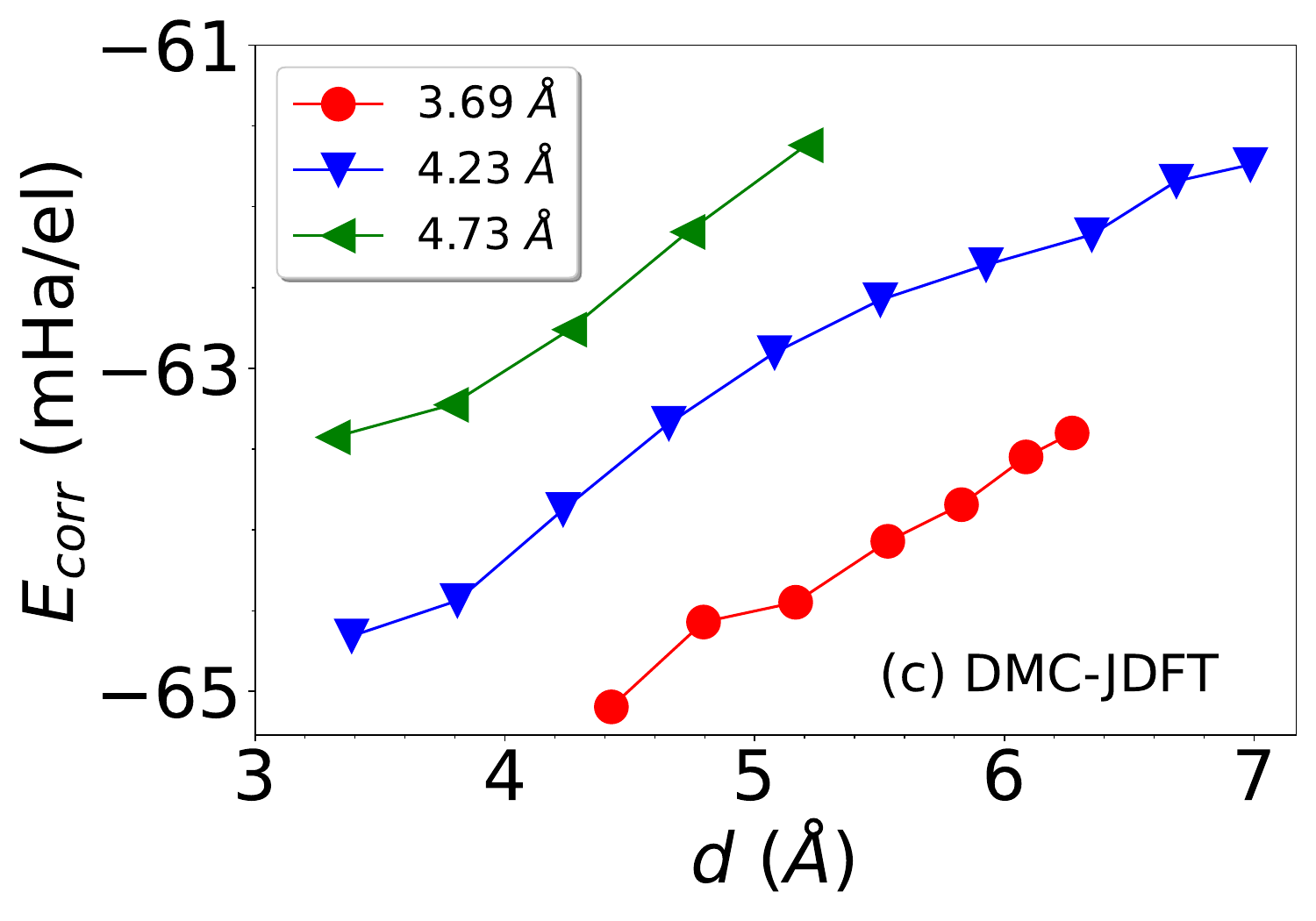}&
    \includegraphics[width=0.45\linewidth]{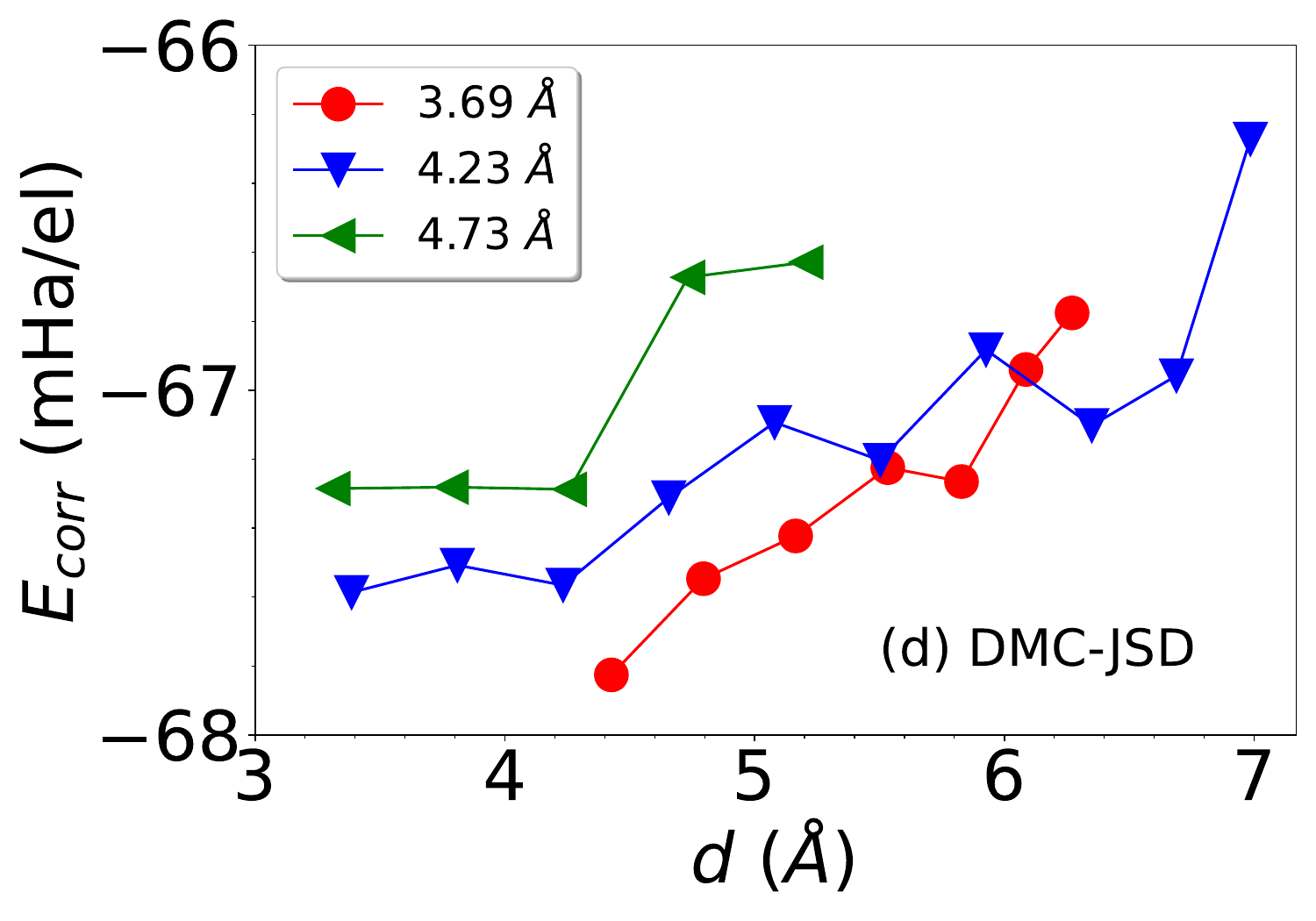}\\
    \end{tabular}
    \caption{Correlation energy per electron for double layer FeSe as a function of interlayer separation obtained using (a) VMC-JDFT, (b) VMC-JSD, (c) DMC-JDFT, and (d) DMC-JSD  for three Fe-Fe bond-length of 3.69, 4.23, and 4.73~\AA.}
    \label{fig:Ecor}
\end{figure}

Figure~\ref{fig:Ecor} presents the VMC and DMC correlation energies of double-layer FeSe as a function of the interlayer distance. The calculations were performed using JDFT and JSD trial wave functions in a simulation cell containing $N=176$ electrons. Both VMC and DMC results consistently show that the magnitude of the correlation energy decreases with increasing interlayer separation. Moreover, JDFT-based wave functions predict that, for a fixed interlayer spacing $d$, the absolute value of $E_{corr}$ decreases with in-plane stretching. The JSD-based VMC and DMC calculations reveal a similar trend, though with smaller variations in $E_{corr}$ across the studied Fe–Fe distances.

Defining the parameter $\Delta_{corr} = (E_{corr}^{Fe} + E_{corr}^{Se})/2 - E_{corr}^{FeSe}$, we find that $\Delta_{corr}$ decreases with increasing interlayer separation as well as with in-plane stretching. This behavior reflects the fact that, upon dissociation of double-layer FeSe into its atomic fragments, $\Delta_{corr} \rightarrow 0$. The rate of change in $\Delta_{corr}$ depends on the wave function and also on the QMC method. 

\section{Conclusion}
We computed the VMC and DMC energies of isolated Fe and Se atoms, as well as of double-layer FeSe, using both JDFT and JSD wave functions. The $E_{\text{corr}}$ obtained for the atoms are in good agreement with the all-electron QC results. For double-layer FeSe, we performed calculations at three different system sizes and extrapolated the energies to the thermodynamic limit to estimate $E_{\text{corr}}$ at infinite system size. Our results indicate that the atomic contributions dominate the correlation energy of double-layer FeSe, while the contribution from bonding is comparatively minor. 

We calculated $E_{corr}$ of double-layer FeSe as a function of in-plane tensile strain and interlayer separation. The results reveal a clear trend that the magnitude of $E_{corr}$ decreases with increasing in-plane stretch and interlayer distance, approaching the values corresponding to isolated atomic fragments \cite{Suppl}.

Our QMC results indicate that in FeSe and potentially in other transition metal compounds, correlation effects remain large and are not reduced relative to the atomic case. In contrast, for simple metals, where the electron liquid model provides a good description, electron delocalization and screening of the Coulomb interaction generally suppress the correlation energy compared to their atomic counterparts. However, in FeSe and related iron-based systems, the Fe-$3d$ states are spatially localized and exhibit narrow bandwidths relative to the $s$ or $p$ states in simple metals. This localization enhances the el–el Coulomb interactions and consequently increases the correlation energy per electron. An obvious conclusion is that in transition metal compounds, the bands are narrower, so the on-site Coulomb repulsion $U$ is more significant relative to the band-width $W$ making the effective correlation strength $U/W$ larger than in simple metals.

\section*{Acknowledgments}
We acknowledge the support of the Leverhulme Trust under the grant agreement RPG-2023-253. S. Azadi and T.D. K\"{u}hne acknowledge the computing time provided to them on the high-performance computers Noctua2 at the NHR Center in Paderborn (PC2).

\bibliography{mainbib}

\begin{thebibliography}{61}%
\makeatletter
\providecommand \@ifxundefined [1]{%
 \@ifx{#1\undefined}
}%
\providecommand \@ifnum [1]{%
 \ifnum #1\expandafter \@firstoftwo
 \else \expandafter \@secondoftwo
 \fi
}%
\providecommand \@ifx [1]{%
 \ifx #1\expandafter \@firstoftwo
 \else \expandafter \@secondoftwo
 \fi
}%
\providecommand \natexlab [1]{#1}%
\providecommand \enquote  [1]{``#1''}%
\providecommand \bibnamefont  [1]{#1}%
\providecommand \bibfnamefont [1]{#1}%
\providecommand \citenamefont [1]{#1}%
\providecommand \href@noop [0]{\@secondoftwo}%
\providecommand \href [0]{\begingroup \@sanitize@url \@href}%
\providecommand \@href[1]{\@@startlink{#1}\@@href}%
\providecommand \@@href[1]{\endgroup#1\@@endlink}%
\providecommand \@sanitize@url [0]{\catcode `\\12\catcode `\$12\catcode
  `\&12\catcode `\#12\catcode `\^12\catcode `\_12\catcode `\%12\relax}%
\providecommand \@@startlink[1]{}%
\providecommand \@@endlink[0]{}%
\providecommand \url  [0]{\begingroup\@sanitize@url \@url }%
\providecommand \@url [1]{\endgroup\@href {#1}{\urlprefix }}%
\providecommand \urlprefix  [0]{URL }%
\providecommand \Eprint [0]{\href }%
\providecommand \doibase [0]{https://doi.org/}%
\providecommand \selectlanguage [0]{\@gobble}%
\providecommand \bibinfo  [0]{\@secondoftwo}%
\providecommand \bibfield  [0]{\@secondoftwo}%
\providecommand \translation [1]{[#1]}%
\providecommand \BibitemOpen [0]{}%
\providecommand \bibitemStop [0]{}%
\providecommand \bibitemNoStop [0]{.\EOS\space}%
\providecommand \EOS [0]{\spacefactor3000\relax}%
\providecommand \BibitemShut  [1]{\csname bibitem#1\endcsname}%
\let\auto@bib@innerbib\@empty
\bibitem [{\citenamefont {Wang}\ \emph {et~al.}(2022)\citenamefont {Wang},
  \citenamefont {Bedoya-Pinto}, \citenamefont {Blei}, \citenamefont {Dismukes},
  \citenamefont {Hamo}, \citenamefont {Jenkins}, \citenamefont {Koperski},
  \citenamefont {Liu}, \citenamefont {Sun}, \citenamefont {Telford},
  \citenamefont {Kim}, \citenamefont {Augustin}, \citenamefont {Vool},
  \citenamefont {Yin}, \citenamefont {Li}, \citenamefont {Falin}, \citenamefont
  {Dean}, \citenamefont {Casanova}, \citenamefont {Evans}, \citenamefont
  {Chshiev}, \citenamefont {Mishchenko}, \citenamefont {Petrovic},
  \citenamefont {He}, \citenamefont {Zhao}, \citenamefont {Tsen}, \citenamefont
  {Gerardot}, \citenamefont {Brotons-Gisbert}, \citenamefont {Guguchia},
  \citenamefont {Roy}, \citenamefont {Tongay}, \citenamefont {Wang},
  \citenamefont {Hasan}, \citenamefont {Wrachtrup}, \citenamefont {Yacoby},
  \citenamefont {Fert}, \citenamefont {Parkin}, \citenamefont {Novoselov},
  \citenamefont {Dai}, \citenamefont {Balicas},\ and\ \citenamefont
  {Santos}}]{HuaWang2022}%
  \BibitemOpen
  \bibfield  {author} {\bibinfo {author} {\bibfnamefont {Q.~H.}\ \bibnamefont
  {Wang}}, \bibinfo {author} {\bibfnamefont {A.}~\bibnamefont {Bedoya-Pinto}},
  \bibinfo {author} {\bibfnamefont {M.}~\bibnamefont {Blei}}, \bibinfo {author}
  {\bibfnamefont {A.~H.}\ \bibnamefont {Dismukes}}, \bibinfo {author}
  {\bibfnamefont {A.}~\bibnamefont {Hamo}}, \bibinfo {author} {\bibfnamefont
  {S.}~\bibnamefont {Jenkins}}, \bibinfo {author} {\bibfnamefont
  {M.}~\bibnamefont {Koperski}}, \bibinfo {author} {\bibfnamefont
  {Y.}~\bibnamefont {Liu}}, \bibinfo {author} {\bibfnamefont {Q.-C.}\
  \bibnamefont {Sun}}, \bibinfo {author} {\bibfnamefont {E.~J.}\ \bibnamefont
  {Telford}}, \bibinfo {author} {\bibfnamefont {H.~H.}\ \bibnamefont {Kim}},
  \bibinfo {author} {\bibfnamefont {M.}~\bibnamefont {Augustin}}, \bibinfo
  {author} {\bibfnamefont {U.}~\bibnamefont {Vool}}, \bibinfo {author}
  {\bibfnamefont {J.-X.}\ \bibnamefont {Yin}}, \bibinfo {author} {\bibfnamefont
  {L.~H.}\ \bibnamefont {Li}}, \bibinfo {author} {\bibfnamefont
  {A.}~\bibnamefont {Falin}}, \bibinfo {author} {\bibfnamefont {C.~R.}\
  \bibnamefont {Dean}}, \bibinfo {author} {\bibfnamefont {F.}~\bibnamefont
  {Casanova}}, \bibinfo {author} {\bibfnamefont {R.~F.}\ \bibnamefont {Evans}},
  \bibinfo {author} {\bibfnamefont {M.}~\bibnamefont {Chshiev}}, \bibinfo
  {author} {\bibfnamefont {A.}~\bibnamefont {Mishchenko}}, \bibinfo {author}
  {\bibfnamefont {C.}~\bibnamefont {Petrovic}}, \bibinfo {author}
  {\bibfnamefont {R.}~\bibnamefont {He}}, \bibinfo {author} {\bibfnamefont
  {L.}~\bibnamefont {Zhao}}, \bibinfo {author} {\bibfnamefont {A.~W.}\
  \bibnamefont {Tsen}}, \bibinfo {author} {\bibfnamefont {B.~D.}\ \bibnamefont
  {Gerardot}}, \bibinfo {author} {\bibfnamefont {M.}~\bibnamefont
  {Brotons-Gisbert}}, \bibinfo {author} {\bibfnamefont {Z.}~\bibnamefont
  {Guguchia}}, \bibinfo {author} {\bibfnamefont {X.}~\bibnamefont {Roy}},
  \bibinfo {author} {\bibfnamefont {S.}~\bibnamefont {Tongay}}, \bibinfo
  {author} {\bibfnamefont {Z.}~\bibnamefont {Wang}}, \bibinfo {author}
  {\bibfnamefont {M.~Z.}\ \bibnamefont {Hasan}}, \bibinfo {author}
  {\bibfnamefont {J.}~\bibnamefont {Wrachtrup}}, \bibinfo {author}
  {\bibfnamefont {A.}~\bibnamefont {Yacoby}}, \bibinfo {author} {\bibfnamefont
  {A.}~\bibnamefont {Fert}}, \bibinfo {author} {\bibfnamefont {S.}~\bibnamefont
  {Parkin}}, \bibinfo {author} {\bibfnamefont {K.~S.}\ \bibnamefont
  {Novoselov}}, \bibinfo {author} {\bibfnamefont {P.}~\bibnamefont {Dai}},
  \bibinfo {author} {\bibfnamefont {L.}~\bibnamefont {Balicas}},\ and\ \bibinfo
  {author} {\bibfnamefont {E.~J.}\ \bibnamefont {Santos}},\ }\bibfield  {title}
  {\bibinfo {title} {The magnetic genome of two-dimensional van der waals
  materials},\ }\href@noop {} {\bibfield  {journal} {\bibinfo  {journal} {ACS
  Nano}\ }\textbf {\bibinfo {volume} {16}},\ \bibinfo {pages} {6960} (\bibinfo
  {year} {2022})}\BibitemShut {NoStop}%
\bibitem [{\citenamefont {Gibertini}\ \emph {et~al.}(2019)\citenamefont
  {Gibertini}, \citenamefont {Koperski}, \citenamefont {Morpurgo},\ and\
  \citenamefont {Novoselov}}]{Gibertini2019}%
  \BibitemOpen
  \bibfield  {author} {\bibinfo {author} {\bibfnamefont {M.}~\bibnamefont
  {Gibertini}}, \bibinfo {author} {\bibfnamefont {M.}~\bibnamefont {Koperski}},
  \bibinfo {author} {\bibfnamefont {A.}~\bibnamefont {Morpurgo}},\ and\
  \bibinfo {author} {\bibfnamefont {K.}~\bibnamefont {Novoselov}},\ }\bibfield
  {title} {\bibinfo {title} {Magnetic 2d materials and heterostructures},\
  }\href@noop {} {\bibfield  {journal} {\bibinfo  {journal} {Nature
  nanotechnology}\ }\textbf {\bibinfo {volume} {14}},\ \bibinfo {pages} {408}
  (\bibinfo {year} {2019})}\BibitemShut {NoStop}%
\bibitem [{\citenamefont {Jenkins}\ \emph {et~al.}(2022)\citenamefont
  {Jenkins}, \citenamefont {R\'{o}zsa}, \citenamefont {Atxitia}, \citenamefont
  {Evans}, \citenamefont {Novoselov},\ and\ \citenamefont
  {Santos}}]{Jenkins2022}%
  \BibitemOpen
  \bibfield  {author} {\bibinfo {author} {\bibfnamefont {S.}~\bibnamefont
  {Jenkins}}, \bibinfo {author} {\bibfnamefont {L.}~\bibnamefont {R\'{o}zsa}},
  \bibinfo {author} {\bibfnamefont {U.}~\bibnamefont {Atxitia}}, \bibinfo
  {author} {\bibfnamefont {R.~F.~L.}\ \bibnamefont {Evans}}, \bibinfo {author}
  {\bibfnamefont {K.~S.}\ \bibnamefont {Novoselov}},\ and\ \bibinfo {author}
  {\bibfnamefont {E.~J.~G.}\ \bibnamefont {Santos}},\ }\bibfield  {title}
  {\bibinfo {title} {Breaking through the mermin-wagner limit in \text{2D} van
  der waals magnets},\ }\href@noop {} {\bibfield  {journal} {\bibinfo
  {journal} {Nat. Comm.}\ }\textbf {\bibinfo {volume} {13}},\ \bibinfo {pages}
  {6917} (\bibinfo {year} {2022})}\BibitemShut {NoStop}%
\bibitem [{\citenamefont {Gong}\ and\ \citenamefont {et~al}(2017)}]{Gong2017}%
  \BibitemOpen
  \bibfield  {author} {\bibinfo {author} {\bibfnamefont {C.}~\bibnamefont
  {Gong}}\ and\ \bibinfo {author} {\bibnamefont {et~al}},\ }\bibfield  {title}
  {\bibinfo {title} {Discovery of intrinsic ferromagnetism in two dimensional
  van der waals crystals},\ }\href@noop {} {\bibfield  {journal} {\bibinfo
  {journal} {Nature}\ }\textbf {\bibinfo {volume} {546}},\ \bibinfo {pages}
  {265} (\bibinfo {year} {2017})}\BibitemShut {NoStop}%
\bibitem [{\citenamefont {Kontani}\ and\ \citenamefont
  {Onari}(2010)}]{Kontani2010}%
  \BibitemOpen
  \bibfield  {author} {\bibinfo {author} {\bibfnamefont {H.}~\bibnamefont
  {Kontani}}\ and\ \bibinfo {author} {\bibfnamefont {S.}~\bibnamefont
  {Onari}},\ }\bibfield  {title} {\bibinfo {title}
  {Orbital-fluctuation-mediated superconductivity in iron pnictides: Analysis
  of the five-orbital hubbard-holstein model},\ }\href@noop {} {\bibfield
  {journal} {\bibinfo  {journal} {Phys. Rev. Lett.}\ }\textbf {\bibinfo
  {volume} {104}},\ \bibinfo {pages} {157001} (\bibinfo {year}
  {2010})}\BibitemShut {NoStop}%
\bibitem [{\citenamefont {B\"{o}hmer}\ and\ \citenamefont
  {Kreisel}(2018)}]{Bohmer2018}%
  \BibitemOpen
  \bibfield  {author} {\bibinfo {author} {\bibfnamefont {A.~E.}\ \bibnamefont
  {B\"{o}hmer}}\ and\ \bibinfo {author} {\bibfnamefont {A.}~\bibnamefont
  {Kreisel}},\ }\bibfield  {title} {\bibinfo {title} {Nematicity, magnetism and
  superconductivity in \text{FeSe}},\ }\href@noop {} {\bibfield  {journal}
  {\bibinfo  {journal} {J. Phys.: Condensed Matter}\ }\textbf {\bibinfo
  {volume} {30}},\ \bibinfo {pages} {023001} (\bibinfo {year}
  {2018})}\BibitemShut {NoStop}%
\bibitem [{\citenamefont {Medvedev}\ \emph {et~al.}(2009)\citenamefont
  {Medvedev}, \citenamefont {McQueen}, \citenamefont {Troyan}, \citenamefont
  {Palasyuk}, \citenamefont {Eremets}, \citenamefont {Cava}, \citenamefont
  {Naghavi}, \citenamefont {Casper}, \citenamefont {Ksenofontov}, \citenamefont
  {Wortmann},\ and\ \citenamefont {Felser}}]{Medvedev2009}%
  \BibitemOpen
  \bibfield  {author} {\bibinfo {author} {\bibfnamefont {S.}~\bibnamefont
  {Medvedev}}, \bibinfo {author} {\bibfnamefont {T.~M.}\ \bibnamefont
  {McQueen}}, \bibinfo {author} {\bibfnamefont {I.~A.}\ \bibnamefont {Troyan}},
  \bibinfo {author} {\bibfnamefont {T.}~\bibnamefont {Palasyuk}}, \bibinfo
  {author} {\bibfnamefont {M.~I.}\ \bibnamefont {Eremets}}, \bibinfo {author}
  {\bibfnamefont {R.~J.}\ \bibnamefont {Cava}}, \bibinfo {author}
  {\bibfnamefont {S.}~\bibnamefont {Naghavi}}, \bibinfo {author} {\bibfnamefont
  {F.}~\bibnamefont {Casper}}, \bibinfo {author} {\bibfnamefont
  {V.}~\bibnamefont {Ksenofontov}}, \bibinfo {author} {\bibfnamefont
  {G.}~\bibnamefont {Wortmann}},\ and\ \bibinfo {author} {\bibfnamefont
  {C.}~\bibnamefont {Felser}},\ }\bibfield  {title} {\bibinfo {title}
  {Electronic and magnetic phase diagram of $\beta$-\text{Fe}$_{1.01}$\text{Se}
  with superconductivity at 36.7 \text{K} under pressure},\ }\href@noop {}
  {\bibfield  {journal} {\bibinfo  {journal} {Nat. Mater.}\ }\textbf {\bibinfo
  {volume} {8}},\ \bibinfo {pages} {630} (\bibinfo {year} {2009})}\BibitemShut
  {NoStop}%
\bibitem [{\citenamefont {Kang}\ \emph {et~al.}(2024)\citenamefont {Kang},
  \citenamefont {Kim}, \citenamefont {Park},\ and\ \citenamefont
  {Janotti}}]{Kang2024}%
  \BibitemOpen
  \bibfield  {author} {\bibinfo {author} {\bibfnamefont {B.}~\bibnamefont
  {Kang}}, \bibinfo {author} {\bibfnamefont {M.}~\bibnamefont {Kim}}, \bibinfo
  {author} {\bibfnamefont {C.~H.}\ \bibnamefont {Park}},\ and\ \bibinfo
  {author} {\bibfnamefont {A.}~\bibnamefont {Janotti}},\ }\bibfield  {title}
  {\bibinfo {title} {Mott-insulator state of \text{FeSe} as a van der waals
  \text{2D} material is unveiled},\ }\href@noop {} {\bibfield  {journal}
  {\bibinfo  {journal} {Phys. Rev. Lett.}\ }\textbf {\bibinfo {volume} {132}},\
  \bibinfo {pages} {266506} (\bibinfo {year} {2024})}\BibitemShut {NoStop}%
\bibitem [{\citenamefont {Dagotto}(2013)}]{Dagotto2013}%
  \BibitemOpen
  \bibfield  {author} {\bibinfo {author} {\bibfnamefont {E.}~\bibnamefont
  {Dagotto}},\ }\bibfield  {title} {\bibinfo {title} {Colloquium: The
  unexpected properties of alkali metal iron selenide superconductors},\
  }\href@noop {} {\bibfield  {journal} {\bibinfo  {journal} {Rev. Mod. Phys.}\
  }\textbf {\bibinfo {volume} {85}},\ \bibinfo {pages} {849} (\bibinfo {year}
  {2013})}\BibitemShut {NoStop}%
\bibitem [{\citenamefont {Dai}(2015)}]{Dai2015}%
  \BibitemOpen
  \bibfield  {author} {\bibinfo {author} {\bibfnamefont {P.}~\bibnamefont
  {Dai}},\ }\bibfield  {title} {\bibinfo {title} {Antiferromagnetic order and
  spin dynamics in iron-based superconductors},\ }\href@noop {} {\bibfield
  {journal} {\bibinfo  {journal} {Rev. Mod. Phys.}\ }\textbf {\bibinfo {volume}
  {87}},\ \bibinfo {pages} {855} (\bibinfo {year} {2015})}\BibitemShut
  {NoStop}%
\bibitem [{\citenamefont {Huang}\ \emph {et~al.}(2020)\citenamefont {Huang},
  \citenamefont {McGuire}, \citenamefont {May}, \citenamefont {Xiao},
  \citenamefont {Jarillo-Herrero},\ and\ \citenamefont {Xu}}]{Huang2020}%
  \BibitemOpen
  \bibfield  {author} {\bibinfo {author} {\bibfnamefont {B.}~\bibnamefont
  {Huang}}, \bibinfo {author} {\bibfnamefont {M.~A.}\ \bibnamefont {McGuire}},
  \bibinfo {author} {\bibfnamefont {A.~F.}\ \bibnamefont {May}}, \bibinfo
  {author} {\bibfnamefont {D.}~\bibnamefont {Xiao}}, \bibinfo {author}
  {\bibfnamefont {P.}~\bibnamefont {Jarillo-Herrero}},\ and\ \bibinfo {author}
  {\bibfnamefont {X.}~\bibnamefont {Xu}},\ }\bibfield  {title} {\bibinfo
  {title} {Emergent phenomena and proximity effects in two-dimensional magnets
  and heterostructures},\ }\href@noop {} {\bibfield  {journal} {\bibinfo
  {journal} {Nature Materials}\ }\textbf {\bibinfo {volume} {19}},\ \bibinfo
  {pages} {1276} (\bibinfo {year} {2020})}\BibitemShut {NoStop}%
\bibitem [{\citenamefont {Kothapalli}\ \emph {et~al.}(2016)\citenamefont
  {Kothapalli}, \citenamefont {B\"{o}hmer}, \citenamefont {Jayasekara},
  \citenamefont {Ueland}, \citenamefont {Das}, \citenamefont {Taufour},
  \citenamefont {Xiao}, \citenamefont {Alp}, \citenamefont {Bud'ko},
  \citenamefont {Canfield}, \citenamefont {Kreyssig},\ and\ \citenamefont
  {Goldman}}]{Kothapalli2016}%
  \BibitemOpen
  \bibfield  {author} {\bibinfo {author} {\bibfnamefont {K.}~\bibnamefont
  {Kothapalli}}, \bibinfo {author} {\bibfnamefont {A.}~\bibnamefont
  {B\"{o}hmer}}, \bibinfo {author} {\bibfnamefont {W.}~\bibnamefont
  {Jayasekara}}, \bibinfo {author} {\bibfnamefont {B.}~\bibnamefont {Ueland}},
  \bibinfo {author} {\bibfnamefont {P.}~\bibnamefont {Das}}, \bibinfo {author}
  {\bibfnamefont {A.~S.~V.}\ \bibnamefont {Taufour}}, \bibinfo {author}
  {\bibfnamefont {Y.}~\bibnamefont {Xiao}}, \bibinfo {author} {\bibfnamefont
  {E.}~\bibnamefont {Alp}}, \bibinfo {author} {\bibfnamefont {S.}~\bibnamefont
  {Bud'ko}}, \bibinfo {author} {\bibfnamefont {P.}~\bibnamefont {Canfield}},
  \bibinfo {author} {\bibfnamefont {A.}~\bibnamefont {Kreyssig}},\ and\
  \bibinfo {author} {\bibfnamefont {A.}~\bibnamefont {Goldman}},\ }\bibfield
  {title} {\bibinfo {title} {Strong cooperative coupling of pressure-induced
  magnetic order and nematicity in \text{FeSe}},\ }\href@noop {} {\bibfield
  {journal} {\bibinfo  {journal} {Nat. Commun. 7:12728}\ }\textbf {\bibinfo
  {volume} {7}},\ \bibinfo {pages} {12728} (\bibinfo {year}
  {2016})}\BibitemShut {NoStop}%
\bibitem [{\citenamefont {Huang}\ \emph {et~al.}(2017)\citenamefont {Huang},
  \citenamefont {Clark}, \citenamefont {Navarro-Moratalla}, \citenamefont
  {Klein}, \citenamefont {Cheng}, \citenamefont {Seyler}, \citenamefont
  {Zhong}, \citenamefont {Schmidgall}, \citenamefont {McGuire}, \citenamefont
  {Cobden}, \citenamefont {Yao}, \citenamefont {Xiao}, \citenamefont
  {Jarillo-Herrero},\ and\ \citenamefont {Xu}}]{Huang2017}%
  \BibitemOpen
  \bibfield  {author} {\bibinfo {author} {\bibfnamefont {B.}~\bibnamefont
  {Huang}}, \bibinfo {author} {\bibfnamefont {G.}~\bibnamefont {Clark}},
  \bibinfo {author} {\bibfnamefont {E.}~\bibnamefont {Navarro-Moratalla}},
  \bibinfo {author} {\bibfnamefont {D.~R.}\ \bibnamefont {Klein}}, \bibinfo
  {author} {\bibfnamefont {R.}~\bibnamefont {Cheng}}, \bibinfo {author}
  {\bibfnamefont {K.~L.}\ \bibnamefont {Seyler}}, \bibinfo {author}
  {\bibfnamefont {D.}~\bibnamefont {Zhong}}, \bibinfo {author} {\bibfnamefont
  {E.}~\bibnamefont {Schmidgall}}, \bibinfo {author} {\bibfnamefont {M.~A.}\
  \bibnamefont {McGuire}}, \bibinfo {author} {\bibfnamefont {D.~H.}\
  \bibnamefont {Cobden}}, \bibinfo {author} {\bibfnamefont {W.}~\bibnamefont
  {Yao}}, \bibinfo {author} {\bibfnamefont {D.}~\bibnamefont {Xiao}}, \bibinfo
  {author} {\bibfnamefont {P.}~\bibnamefont {Jarillo-Herrero}},\ and\ \bibinfo
  {author} {\bibfnamefont {X.}~\bibnamefont {Xu}},\ }\bibfield  {title}
  {\bibinfo {title} {Layer-dependent ferromagnetism in a van der waals crystal
  down to the monolayer limit},\ }\href@noop {} {\bibfield  {journal} {\bibinfo
   {journal} {Nature}\ }\textbf {\bibinfo {volume} {546}},\ \bibinfo {pages}
  {270} (\bibinfo {year} {2017})}\BibitemShut {NoStop}%
\bibitem [{\citenamefont {Zhang}\ \emph {et~al.}(2022)\citenamefont {Zhang},
  \citenamefont {Pincelli}, \citenamefont {Jozwiak}, \citenamefont {Kondo},
  \citenamefont {Ernstorfer}, \citenamefont {Sato},\ and\ \citenamefont
  {Zhou}}]{Zhang2022}%
  \BibitemOpen
  \bibfield  {author} {\bibinfo {author} {\bibfnamefont {H.}~\bibnamefont
  {Zhang}}, \bibinfo {author} {\bibfnamefont {T.}~\bibnamefont {Pincelli}},
  \bibinfo {author} {\bibfnamefont {C.}~\bibnamefont {Jozwiak}}, \bibinfo
  {author} {\bibfnamefont {T.}~\bibnamefont {Kondo}}, \bibinfo {author}
  {\bibfnamefont {R.}~\bibnamefont {Ernstorfer}}, \bibinfo {author}
  {\bibfnamefont {T.}~\bibnamefont {Sato}},\ and\ \bibinfo {author}
  {\bibfnamefont {S.}~\bibnamefont {Zhou}},\ }\bibfield  {title} {\bibinfo
  {title} {Angle-resolved photoemission spectroscopy},\ }\href@noop {}
  {\bibfield  {journal} {\bibinfo  {journal} {Nature Reviews Methods Primers}\
  }\textbf {\bibinfo {volume} {2}},\ \bibinfo {pages} {54} (\bibinfo {year}
  {2022})}\BibitemShut {NoStop}%
\bibitem [{\citenamefont {Landau}(1958)}]{Landau}%
  \BibitemOpen
  \bibfield  {author} {\bibinfo {author} {\bibfnamefont {L.~D.}\ \bibnamefont
  {Landau}},\ }\bibfield  {title} {\bibinfo {title} {On the theory of the fermi
  liquid},\ }\href@noop {} {\bibfield  {journal} {\bibinfo  {journal} {Sov.
  Phys. JETP}\ }\textbf {\bibinfo {volume} {35}},\ \bibinfo {pages} {95}
  (\bibinfo {year} {1958})}\BibitemShut {NoStop}%
\bibitem [{\citenamefont {Tazai}\ \emph {et~al.}(2023)\citenamefont {Tazai},
  \citenamefont {Matsubara}, \citenamefont {Yamakawa}, \citenamefont {Onari},\
  and\ \citenamefont {Kontani}}]{Tazai2023}%
  \BibitemOpen
  \bibfield  {author} {\bibinfo {author} {\bibfnamefont {R.}~\bibnamefont
  {Tazai}}, \bibinfo {author} {\bibfnamefont {S.}~\bibnamefont {Matsubara}},
  \bibinfo {author} {\bibfnamefont {Y.}~\bibnamefont {Yamakawa}}, \bibinfo
  {author} {\bibfnamefont {S.}~\bibnamefont {Onari}},\ and\ \bibinfo {author}
  {\bibfnamefont {H.}~\bibnamefont {Kontani}},\ }\bibfield  {title} {\bibinfo
  {title} {Rigorous formalism for unconventional symmetry breaking in fermi
  liquid theory and its application to nematicity in \text{FeSe}},\ }\href@noop
  {} {\bibfield  {journal} {\bibinfo  {journal} {Phys. Rev. B}\ }\textbf
  {\bibinfo {volume} {107}},\ \bibinfo {pages} {035137} (\bibinfo {year}
  {2023})}\BibitemShut {NoStop}%
\bibitem [{\citenamefont {Chubukov}\ \emph {et~al.}(2016)\citenamefont
  {Chubukov}, \citenamefont {Khodas},\ and\ \citenamefont
  {Fernandes}}]{Chubukov2016}%
  \BibitemOpen
  \bibfield  {author} {\bibinfo {author} {\bibfnamefont {A.~V.}\ \bibnamefont
  {Chubukov}}, \bibinfo {author} {\bibfnamefont {M.}~\bibnamefont {Khodas}},\
  and\ \bibinfo {author} {\bibfnamefont {R.~M.}\ \bibnamefont {Fernandes}},\
  }\bibfield  {title} {\bibinfo {title} {Superconductivity, and spontaneous
  orbital order in iron-based superconductors: Which comes first and why?},\
  }\href@noop {} {\bibfield  {journal} {\bibinfo  {journal} {Phys. Rev. X}\
  }\textbf {\bibinfo {volume} {6}},\ \bibinfo {pages} {041045} (\bibinfo {year}
  {2016})}\BibitemShut {NoStop}%
\bibitem [{\citenamefont {Yamakawa}\ \emph {et~al.}(2016)\citenamefont
  {Yamakawa}, \citenamefont {Onari},\ and\ \citenamefont
  {Kontani}}]{Yamakawa2016}%
  \BibitemOpen
  \bibfield  {author} {\bibinfo {author} {\bibfnamefont {Y.}~\bibnamefont
  {Yamakawa}}, \bibinfo {author} {\bibfnamefont {S.}~\bibnamefont {Onari}},\
  and\ \bibinfo {author} {\bibfnamefont {H.}~\bibnamefont {Kontani}},\
  }\bibfield  {title} {\bibinfo {title} {Nematicity and magnetism in
  \text{FeSe} and other families of \text{Fe}-based superconductors},\
  }\href@noop {} {\bibfield  {journal} {\bibinfo  {journal} {Phys. Rev. X}\
  }\textbf {\bibinfo {volume} {6}},\ \bibinfo {pages} {021032} (\bibinfo {year}
  {2016})}\BibitemShut {NoStop}%
\bibitem [{\citenamefont {Onari}\ \emph {et~al.}(2016)\citenamefont {Onari},
  \citenamefont {Yamakawa},\ and\ \citenamefont {Kontani}}]{Onari2016}%
  \BibitemOpen
  \bibfield  {author} {\bibinfo {author} {\bibfnamefont {S.}~\bibnamefont
  {Onari}}, \bibinfo {author} {\bibfnamefont {Y.}~\bibnamefont {Yamakawa}},\
  and\ \bibinfo {author} {\bibfnamefont {H.}~\bibnamefont {Kontani}},\
  }\bibfield  {title} {\bibinfo {title} {Sign-reversing orbital polarization in
  the nematic phase of \text{FeSe} due to the symmetry breaking in the
  self-energy},\ }\href@noop {} {\bibfield  {journal} {\bibinfo  {journal}
  {Phys. Rev. Lett.}\ }\textbf {\bibinfo {volume} {116}},\ \bibinfo {pages}
  {227001} (\bibinfo {year} {2016})}\BibitemShut {NoStop}%
\bibitem [{\citenamefont {Fernandes}\ \emph {et~al.}(2014)\citenamefont
  {Fernandes}, \citenamefont {Chubukov},\ and\ \citenamefont
  {Schmalian}}]{Fernandes2014}%
  \BibitemOpen
  \bibfield  {author} {\bibinfo {author} {\bibfnamefont {R.}~\bibnamefont
  {Fernandes}}, \bibinfo {author} {\bibfnamefont {A.}~\bibnamefont
  {Chubukov}},\ and\ \bibinfo {author} {\bibfnamefont {J.}~\bibnamefont
  {Schmalian}},\ }\bibfield  {title} {\bibinfo {title} {What drives nematic
  order in iron-based superconductors},\ }\href@noop {} {\bibfield  {journal}
  {\bibinfo  {journal} {Nat. Phys.}\ }\textbf {\bibinfo {volume} {10}},\
  \bibinfo {pages} {97} (\bibinfo {year} {2014})}\BibitemShut {NoStop}%
\bibitem [{\citenamefont {McQueen}\ \emph {et~al.}(2009)\citenamefont
  {McQueen}, \citenamefont {J.Williams}, \citenamefont {Stephens},
  \citenamefont {Tao}, \citenamefont {Zhu}, \citenamefont {Ksenofontov},
  \citenamefont {Casper}, \citenamefont {Felser},\ and\ \citenamefont
  {Cava}}]{McQueen2009}%
  \BibitemOpen
  \bibfield  {author} {\bibinfo {author} {\bibfnamefont {T.~M.}\ \bibnamefont
  {McQueen}}, \bibinfo {author} {\bibfnamefont {A.}~\bibnamefont {J.Williams}},
  \bibinfo {author} {\bibfnamefont {P.}~\bibnamefont {Stephens}}, \bibinfo
  {author} {\bibfnamefont {J.}~\bibnamefont {Tao}}, \bibinfo {author}
  {\bibfnamefont {Y.}~\bibnamefont {Zhu}}, \bibinfo {author} {\bibfnamefont
  {V.}~\bibnamefont {Ksenofontov}}, \bibinfo {author} {\bibfnamefont
  {F.}~\bibnamefont {Casper}}, \bibinfo {author} {\bibfnamefont
  {C.}~\bibnamefont {Felser}},\ and\ \bibinfo {author} {\bibfnamefont {R.~J.}\
  \bibnamefont {Cava}},\ }\bibfield  {title} {\bibinfo {title}
  {Tetragonal-to-orthorhombic structural phase transition at \text{90K} in the
  superconductor \text{Fe}$_{1.01}$\text{Se}},\ }\href@noop {} {\bibfield
  {journal} {\bibinfo  {journal} {Phys. Rev. Lett.}\ }\textbf {\bibinfo
  {volume} {103}},\ \bibinfo {pages} {057002} (\bibinfo {year}
  {2009})}\BibitemShut {NoStop}%
\bibitem [{\citenamefont {Alloul}\ \emph {et~al.}(2009)\citenamefont {Alloul},
  \citenamefont {Bobroff}, \citenamefont {Gabay},\ and\ \citenamefont
  {Hirschfeld}}]{Alloul2009}%
  \BibitemOpen
  \bibfield  {author} {\bibinfo {author} {\bibfnamefont {H.}~\bibnamefont
  {Alloul}}, \bibinfo {author} {\bibfnamefont {J.}~\bibnamefont {Bobroff}},
  \bibinfo {author} {\bibfnamefont {M.}~\bibnamefont {Gabay}},\ and\ \bibinfo
  {author} {\bibfnamefont {P.~J.}\ \bibnamefont {Hirschfeld}},\ }\bibfield
  {title} {\bibinfo {title} {Defects in correlated metals and
  superconductors},\ }\href@noop {} {\bibfield  {journal} {\bibinfo  {journal}
  {Rev. Mod. Phys.}\ }\textbf {\bibinfo {volume} {81}},\ \bibinfo {pages} {45}
  (\bibinfo {year} {2009})}\BibitemShut {NoStop}%
\bibitem [{\citenamefont {Yi}\ \emph {et~al.}(2013)\citenamefont {Yi},
  \citenamefont {Lu}, \citenamefont {Yu}, \citenamefont {Riggs}, \citenamefont
  {Chu}, \citenamefont {Lv}, \citenamefont {Liu}, \citenamefont {Lu},
  \citenamefont {Cui}, \citenamefont {Hashimoto}, \citenamefont {Mo},
  \citenamefont {Hussain}, \citenamefont {Chu}, \citenamefont {Fisher},
  \citenamefont {Si},\ and\ \citenamefont {Shen}}]{MYi2013}%
  \BibitemOpen
  \bibfield  {author} {\bibinfo {author} {\bibfnamefont {M.}~\bibnamefont
  {Yi}}, \bibinfo {author} {\bibfnamefont {D.~H.}\ \bibnamefont {Lu}}, \bibinfo
  {author} {\bibfnamefont {R.}~\bibnamefont {Yu}}, \bibinfo {author}
  {\bibfnamefont {S.~C.}\ \bibnamefont {Riggs}}, \bibinfo {author}
  {\bibfnamefont {J.-H.}\ \bibnamefont {Chu}}, \bibinfo {author} {\bibfnamefont
  {B.}~\bibnamefont {Lv}}, \bibinfo {author} {\bibfnamefont {Z.~K.}\
  \bibnamefont {Liu}}, \bibinfo {author} {\bibfnamefont {M.}~\bibnamefont
  {Lu}}, \bibinfo {author} {\bibfnamefont {Y.-T.}\ \bibnamefont {Cui}},
  \bibinfo {author} {\bibfnamefont {M.}~\bibnamefont {Hashimoto}}, \bibinfo
  {author} {\bibfnamefont {S.-K.}\ \bibnamefont {Mo}}, \bibinfo {author}
  {\bibfnamefont {Z.}~\bibnamefont {Hussain}}, \bibinfo {author} {\bibfnamefont
  {C.}~\bibnamefont {Chu}}, \bibinfo {author} {\bibfnamefont {I.~R.}\
  \bibnamefont {Fisher}}, \bibinfo {author} {\bibfnamefont {Q.}~\bibnamefont
  {Si}},\ and\ \bibinfo {author} {\bibfnamefont {Z.-X.}\ \bibnamefont {Shen}},\
  }\bibfield  {title} {\bibinfo {title} {Observation of temperature-induced
  crossover to an orbital-selective mott phase in
  \text{A}\text{Fe}$_{2-y}$\text{Se}$_2$ (\text{A}=\text{K}, \text{Rb})
  superconductors},\ }\href@noop {} {\bibfield  {journal} {\bibinfo  {journal}
  {Phys. Rev. Lett.}\ }\textbf {\bibinfo {volume} {110}},\ \bibinfo {pages}
  {067003} (\bibinfo {year} {2013})}\BibitemShut {NoStop}%
\bibitem [{\citenamefont {Yin}\ \emph {et~al.}(2011)\citenamefont {Yin},
  \citenamefont {Haule},\ and\ \citenamefont {Kotliar}}]{ZYin2011}%
  \BibitemOpen
  \bibfield  {author} {\bibinfo {author} {\bibfnamefont {Z.}~\bibnamefont
  {Yin}}, \bibinfo {author} {\bibfnamefont {K.}~\bibnamefont {Haule}},\ and\
  \bibinfo {author} {\bibfnamefont {G.}~\bibnamefont {Kotliar}},\ }\bibfield
  {title} {\bibinfo {title} {Kinetic frustration and the nature of the magnetic
  and paramagnetic states in iron pnictides and iron chalcogenides},\
  }\href@noop {} {\bibfield  {journal} {\bibinfo  {journal} {Nat. Mater.}\
  }\textbf {\bibinfo {volume} {10}},\ \bibinfo {pages} {932} (\bibinfo {year}
  {2011})}\BibitemShut {NoStop}%
\bibitem [{\citenamefont {Ma}\ \emph {et~al.}(2017)\citenamefont {Ma},
  \citenamefont {Bourges}, \citenamefont {Sidis}, \citenamefont {Xu},
  \citenamefont {Li}, \citenamefont {Hu}, \citenamefont {Li}, \citenamefont
  {Wang},\ and\ \citenamefont {Li}}]{MMa2017}%
  \BibitemOpen
  \bibfield  {author} {\bibinfo {author} {\bibfnamefont {M.}~\bibnamefont
  {Ma}}, \bibinfo {author} {\bibfnamefont {P.}~\bibnamefont {Bourges}},
  \bibinfo {author} {\bibfnamefont {Y.}~\bibnamefont {Sidis}}, \bibinfo
  {author} {\bibfnamefont {Y.}~\bibnamefont {Xu}}, \bibinfo {author}
  {\bibfnamefont {S.}~\bibnamefont {Li}}, \bibinfo {author} {\bibfnamefont
  {B.}~\bibnamefont {Hu}}, \bibinfo {author} {\bibfnamefont {J.}~\bibnamefont
  {Li}}, \bibinfo {author} {\bibfnamefont {F.}~\bibnamefont {Wang}},\ and\
  \bibinfo {author} {\bibfnamefont {Y.}~\bibnamefont {Li}},\ }\bibfield
  {title} {\bibinfo {title} {Prominent role of spin-orbit coupling in
  \text{FeSe} revealed by inelastic neutron scattering},\ }\href@noop {}
  {\bibfield  {journal} {\bibinfo  {journal} {Phys. Rev. X}\ }\textbf {\bibinfo
  {volume} {7}},\ \bibinfo {pages} {021025} (\bibinfo {year}
  {2017})}\BibitemShut {NoStop}%
\bibitem [{\citenamefont {Azadi}\ \emph {et~al.}(2025)\citenamefont {Azadi},
  \citenamefont {Bahramy},\ and\ \citenamefont {K\"{u}hne}}]{Azadi2025}%
  \BibitemOpen
  \bibfield  {author} {\bibinfo {author} {\bibfnamefont {S.}~\bibnamefont
  {Azadi}}, \bibinfo {author} {\bibfnamefont {M.}~\bibnamefont {Bahramy}},\
  and\ \bibinfo {author} {\bibfnamefont {T.}~\bibnamefont {K\"{u}hne}},\
  }\bibfield  {title} {\bibinfo {title} {Electron correlation effects and
  spin-liquid state in the herbertsmithite kagome lattice},\ }\href@noop {}
  {\bibfield  {journal} {\bibinfo  {journal} {Phys. Rev. Res.}\ }\textbf
  {\bibinfo {volume} {7}},\ \bibinfo {pages} {013165} (\bibinfo {year}
  {2025})}\BibitemShut {NoStop}%
\bibitem [{\citenamefont {Hsu}\ \emph {et~al.}(2008)\citenamefont {Hsu},
  \citenamefont {Luo}, \citenamefont {Yeh}, \citenamefont {Chen}, \citenamefont
  {Huang}, \citenamefont {Wu}, \citenamefont {Lee}, \citenamefont {Huang},
  \citenamefont {Chu}, \citenamefont {Yan},\ and\ \citenamefont
  {Wu}}]{Hsu2008}%
  \BibitemOpen
  \bibfield  {author} {\bibinfo {author} {\bibfnamefont {F.-C.}\ \bibnamefont
  {Hsu}}, \bibinfo {author} {\bibfnamefont {J.-Y.}\ \bibnamefont {Luo}},
  \bibinfo {author} {\bibfnamefont {K.-W.}\ \bibnamefont {Yeh}}, \bibinfo
  {author} {\bibfnamefont {T.-K.}\ \bibnamefont {Chen}}, \bibinfo {author}
  {\bibfnamefont {T.-W.}\ \bibnamefont {Huang}}, \bibinfo {author}
  {\bibfnamefont {P.~M.}\ \bibnamefont {Wu}}, \bibinfo {author} {\bibfnamefont
  {Y.-C.}\ \bibnamefont {Lee}}, \bibinfo {author} {\bibfnamefont {Y.-L.}\
  \bibnamefont {Huang}}, \bibinfo {author} {\bibfnamefont {Y.-Y.}\ \bibnamefont
  {Chu}}, \bibinfo {author} {\bibfnamefont {D.-C.}\ \bibnamefont {Yan}},\ and\
  \bibinfo {author} {\bibfnamefont {M.-K.}\ \bibnamefont {Wu}},\ }\bibfield
  {title} {\bibinfo {title} {Superconductivity in the \text{PbO}-type structure
  $\alpha$-\text{FeSe}},\ }\href@noop {} {\bibfield  {journal} {\bibinfo
  {journal} {Proc. Nat. Acad. Sci.}\ }\textbf {\bibinfo {volume} {105}},\
  \bibinfo {pages} {14262} (\bibinfo {year} {2008})}\BibitemShut {NoStop}%
\bibitem [{\citenamefont {Wang}\ \emph {et~al.}(2020)\citenamefont {Wang},
  \citenamefont {Zhao}, \citenamefont {Koch}, \citenamefont {Billinge},\ and\
  \citenamefont {Zunger}}]{Wang2020}%
  \BibitemOpen
  \bibfield  {author} {\bibinfo {author} {\bibfnamefont {Z.}~\bibnamefont
  {Wang}}, \bibinfo {author} {\bibfnamefont {X.-G.}\ \bibnamefont {Zhao}},
  \bibinfo {author} {\bibfnamefont {R.}~\bibnamefont {Koch}}, \bibinfo {author}
  {\bibfnamefont {S.~J.~L.}\ \bibnamefont {Billinge}},\ and\ \bibinfo {author}
  {\bibfnamefont {A.}~\bibnamefont {Zunger}},\ }\bibfield  {title} {\bibinfo
  {title} {Understanding electronic peculiarities in tetragonal \text{FeSe} as
  local structural symmetry breaking},\ }\href@noop {} {\bibfield  {journal}
  {\bibinfo  {journal} {Phys. Rev. B}\ }\textbf {\bibinfo {volume} {102}},\
  \bibinfo {pages} {235121} (\bibinfo {year} {2020})}\BibitemShut {NoStop}%
\bibitem [{\citenamefont {Okabe}\ \emph {et~al.}(2010)\citenamefont {Okabe},
  \citenamefont {Takeshita}, \citenamefont {Horigane}, \citenamefont {T},\ and\
  \citenamefont {Akimitsu}}]{Okabe2010}%
  \BibitemOpen
  \bibfield  {author} {\bibinfo {author} {\bibfnamefont {H.}~\bibnamefont
  {Okabe}}, \bibinfo {author} {\bibfnamefont {N.}~\bibnamefont {Takeshita}},
  \bibinfo {author} {\bibfnamefont {K.}~\bibnamefont {Horigane}}, \bibinfo
  {author} {\bibfnamefont {T.~M.}\ \bibnamefont {T}},\ and\ \bibinfo {author}
  {\bibfnamefont {J.}~\bibnamefont {Akimitsu}},\ }\bibfield  {title} {\bibinfo
  {title} {Pressure-induced high-\text{T}$_c$ superconducting phase in
  \text{FeSe}: correlation between anion height and \text{T}$_c$},\ }\href@noop
  {} {\bibfield  {journal} {\bibinfo  {journal} {Phys. Rev. B}\ }\textbf
  {\bibinfo {volume} {81}},\ \bibinfo {pages} {205119} (\bibinfo {year}
  {2010})}\BibitemShut {NoStop}%
\bibitem [{\citenamefont {Tomczak}\ \emph {et~al.}(2012)\citenamefont
  {Tomczak}, \citenamefont {van Schilfgaarde},\ and\ \citenamefont
  {Kotliar}}]{Tomczak2012}%
  \BibitemOpen
  \bibfield  {author} {\bibinfo {author} {\bibfnamefont {J.~M.}\ \bibnamefont
  {Tomczak}}, \bibinfo {author} {\bibfnamefont {M.}~\bibnamefont {van
  Schilfgaarde}},\ and\ \bibinfo {author} {\bibfnamefont {G.}~\bibnamefont
  {Kotliar}},\ }\bibfield  {title} {\bibinfo {title} {Many-body effects in iron
  pnictides and chalcogenides: Nonlocal versus dynamic origin of effective
  masses},\ }\href@noop {} {\bibfield  {journal} {\bibinfo  {journal} {Phys.
  Rev. Lett.}\ }\textbf {\bibinfo {volume} {109}},\ \bibinfo {pages} {237010}
  (\bibinfo {year} {2012})}\BibitemShut {NoStop}%
\bibitem [{\citenamefont {Long}\ \emph {et~al.}(2020)\citenamefont {Long},
  \citenamefont {Zhang}, \citenamefont {Wang},\ and\ \citenamefont
  {Liu}}]{XLong2020}%
  \BibitemOpen
  \bibfield  {author} {\bibinfo {author} {\bibfnamefont {X.}~\bibnamefont
  {Long}}, \bibinfo {author} {\bibfnamefont {S.}~\bibnamefont {Zhang}},
  \bibinfo {author} {\bibfnamefont {F.}~\bibnamefont {Wang}},\ and\ \bibinfo
  {author} {\bibfnamefont {Z.}~\bibnamefont {Liu}},\ }\bibfield  {title}
  {\bibinfo {title} {A first-principle perspective on electronic nematicity in
  \text{FeSe}},\ }\href@noop {} {\bibfield  {journal} {\bibinfo  {journal} {Npj
  Quantum Mater.}\ }\textbf {\bibinfo {volume} {5}},\ \bibinfo {pages} {50}
  (\bibinfo {year} {2020})}\BibitemShut {NoStop}%
\bibitem [{\citenamefont {Ding}\ \emph {et~al.}(2021)\citenamefont {Ding},
  \citenamefont {Zeng}, \citenamefont {Zheng}, \citenamefont {Zhang},
  \citenamefont {Xu}, \citenamefont {Chen}, \citenamefont {Wang}, \citenamefont
  {Chen}, \citenamefont {Xie}, \citenamefont {Ding}, \citenamefont {Zheng},
  \citenamefont {Zhao}, \citenamefont {Gao},\ and\ \citenamefont
  {Fu}}]{Ding2021}%
  \BibitemOpen
  \bibfield  {author} {\bibinfo {author} {\bibfnamefont {Y.}~\bibnamefont
  {Ding}}, \bibinfo {author} {\bibfnamefont {M.}~\bibnamefont {Zeng}}, \bibinfo
  {author} {\bibfnamefont {Q.}~\bibnamefont {Zheng}}, \bibinfo {author}
  {\bibfnamefont {J.}~\bibnamefont {Zhang}}, \bibinfo {author} {\bibfnamefont
  {D.}~\bibnamefont {Xu}}, \bibinfo {author} {\bibfnamefont {W.}~\bibnamefont
  {Chen}}, \bibinfo {author} {\bibfnamefont {C.}~\bibnamefont {Wang}}, \bibinfo
  {author} {\bibfnamefont {S.}~\bibnamefont {Chen}}, \bibinfo {author}
  {\bibfnamefont {Y.}~\bibnamefont {Xie}}, \bibinfo {author} {\bibfnamefont
  {Y.}~\bibnamefont {Ding}}, \bibinfo {author} {\bibfnamefont {S.}~\bibnamefont
  {Zheng}}, \bibinfo {author} {\bibfnamefont {J.}~\bibnamefont {Zhao}},
  \bibinfo {author} {\bibfnamefont {P.}~\bibnamefont {Gao}},\ and\ \bibinfo
  {author} {\bibfnamefont {L.}~\bibnamefont {Fu}},\ }\bibfield  {title}
  {\bibinfo {title} {Bidirectional and reversible tuning of the interlayer
  spacing of two-dimensional materials},\ }\href@noop {} {\bibfield  {journal}
  {\bibinfo  {journal} {Nat. Comm.}\ }\textbf {\bibinfo {volume} {12}},\
  \bibinfo {pages} {5886} (\bibinfo {year} {2021})}\BibitemShut {NoStop}%
\bibitem [{\citenamefont {Koloren\v{c}}\ \emph {et~al.}(2010)\citenamefont
  {Koloren\v{c}}, \citenamefont {Hu},\ and\ \citenamefont
  {Mitas}}]{Kolorenc2010}%
  \BibitemOpen
  \bibfield  {author} {\bibinfo {author} {\bibfnamefont {J.}~\bibnamefont
  {Koloren\v{c}}}, \bibinfo {author} {\bibfnamefont {S.}~\bibnamefont {Hu}},\
  and\ \bibinfo {author} {\bibfnamefont {L.}~\bibnamefont {Mitas}},\ }\bibfield
   {title} {\bibinfo {title} {Wave functions for quantum monate carlo
  calculations in solids: Orbitals from density functional theory with hybrid
  exchange-correlation functionals},\ }\href@noop {} {\bibfield  {journal}
  {\bibinfo  {journal} {Phys. Rev. B}\ }\textbf {\bibinfo {volume} {82}},\
  \bibinfo {pages} {115108} (\bibinfo {year} {2010})}\BibitemShut {NoStop}%
\bibitem [{\citenamefont {Koloren\v{c}}\ and\ \citenamefont
  {Mitas}(2011)}]{Koloren2011}%
  \BibitemOpen
  \bibfield  {author} {\bibinfo {author} {\bibfnamefont {J.}~\bibnamefont
  {Koloren\v{c}}}\ and\ \bibinfo {author} {\bibfnamefont {L.}~\bibnamefont
  {Mitas}},\ }\bibfield  {title} {\bibinfo {title} {Applications of quantum
  monte carlo methods in condensed systems},\ }\href@noop {} {\bibfield
  {journal} {\bibinfo  {journal} {Reports on Progress in Physics}\ }\textbf
  {\bibinfo {volume} {74}},\ \bibinfo {pages} {026502} (\bibinfo {year}
  {2011})}\BibitemShut {NoStop}%
\bibitem [{\citenamefont {Dubecky}\ \emph {et~al.}(2016)\citenamefont
  {Dubecky}, \citenamefont {Mitas},\ and\ \citenamefont
  {Jurecka}}]{Dubecky2016}%
  \BibitemOpen
  \bibfield  {author} {\bibinfo {author} {\bibfnamefont {M.}~\bibnamefont
  {Dubecky}}, \bibinfo {author} {\bibfnamefont {L.}~\bibnamefont {Mitas}},\
  and\ \bibinfo {author} {\bibfnamefont {P.}~\bibnamefont {Jurecka}},\
  }\bibfield  {title} {\bibinfo {title} {Noncovalent interactions by quantum
  monte carlo},\ }\href@noop {} {\bibfield  {journal} {\bibinfo  {journal}
  {Chemical Reviews}\ }\textbf {\bibinfo {volume} {116}},\ \bibinfo {pages}
  {5188} (\bibinfo {year} {2016})}\BibitemShut {NoStop}%
\bibitem [{\citenamefont {Koloren\v{c}}\ and\ \citenamefont
  {Mitas}(2008)}]{Koloren2008}%
  \BibitemOpen
  \bibfield  {author} {\bibinfo {author} {\bibfnamefont {J.}~\bibnamefont
  {Koloren\v{c}}}\ and\ \bibinfo {author} {\bibfnamefont {L.}~\bibnamefont
  {Mitas}},\ }\bibfield  {title} {\bibinfo {title} {Quantum monte carlo
  calculations of structural properties of \text{FeO} under pressure},\
  }\href@noop {} {\bibfield  {journal} {\bibinfo  {journal} {Phys. Rev. Lett.}\
  }\textbf {\bibinfo {volume} {101}},\ \bibinfo {pages} {185502} (\bibinfo
  {year} {2008})}\BibitemShut {NoStop}%
\bibitem [{\citenamefont {Wines}\ \emph {et~al.}(2025)\citenamefont {Wines},
  \citenamefont {Ahn}, \citenamefont {Benali}, \citenamefont {Kent},
  \citenamefont {Krogel}, \citenamefont {Kwon}, \citenamefont {Mitas},
  \citenamefont {Reboredo}, \citenamefont {Rubenstein}, \citenamefont
  {Saritas}, \citenamefont {Shin}, \citenamefont {\v{S}tich},\ and\
  \citenamefont {Ataca}}]{Wines2025}%
  \BibitemOpen
  \bibfield  {author} {\bibinfo {author} {\bibfnamefont {D.}~\bibnamefont
  {Wines}}, \bibinfo {author} {\bibfnamefont {J.}~\bibnamefont {Ahn}}, \bibinfo
  {author} {\bibfnamefont {A.}~\bibnamefont {Benali}}, \bibinfo {author}
  {\bibfnamefont {P.}~\bibnamefont {Kent}}, \bibinfo {author} {\bibfnamefont
  {J.}~\bibnamefont {Krogel}}, \bibinfo {author} {\bibfnamefont
  {Y.}~\bibnamefont {Kwon}}, \bibinfo {author} {\bibfnamefont {L.}~\bibnamefont
  {Mitas}}, \bibinfo {author} {\bibfnamefont {F.}~\bibnamefont {Reboredo}},
  \bibinfo {author} {\bibfnamefont {B.}~\bibnamefont {Rubenstein}}, \bibinfo
  {author} {\bibfnamefont {K.}~\bibnamefont {Saritas}}, \bibinfo {author}
  {\bibfnamefont {H.}~\bibnamefont {Shin}}, \bibinfo {author} {\bibfnamefont
  {I.}~\bibnamefont {\v{S}tich}},\ and\ \bibinfo {author} {\bibfnamefont
  {C.}~\bibnamefont {Ataca}},\ }\bibfield  {title} {\bibinfo {title} {Toward
  improved property prediction of 2d materials using many-body quantum monte
  carlo methods},\ }\href@noop {} {\bibfield  {journal} {\bibinfo  {journal}
  {Appl. Phys. Rev.}\ }\textbf {\bibinfo {volume} {12}},\ \bibinfo {pages}
  {031317} (\bibinfo {year} {2025})}\BibitemShut {NoStop}%
\bibitem [{\citenamefont {Busemeyer}\ \emph {et~al.}(2016)\citenamefont
  {Busemeyer}, \citenamefont {Dagrada}, \citenamefont {Sorella}, \citenamefont
  {Casula}, ,\ and\ \citenamefont {Wagner}}]{Busemeyer2016}%
  \BibitemOpen
  \bibfield  {author} {\bibinfo {author} {\bibfnamefont {B.}~\bibnamefont
  {Busemeyer}}, \bibinfo {author} {\bibfnamefont {M.}~\bibnamefont {Dagrada}},
  \bibinfo {author} {\bibfnamefont {S.}~\bibnamefont {Sorella}}, \bibinfo
  {author} {\bibfnamefont {M.}~\bibnamefont {Casula}}, ,\ and\ \bibinfo
  {author} {\bibfnamefont {L.~K.}\ \bibnamefont {Wagner}},\ }\bibfield  {title}
  {\bibinfo {title} {Competing collinear magnetic structures in superconducting
  \text{FeSe} by first-principles quantum monte carlo calculations},\
  }\href@noop {} {\bibfield  {journal} {\bibinfo  {journal} {Phys. Rev. B}\
  }\textbf {\bibinfo {volume} {94}},\ \bibinfo {pages} {035108} (\bibinfo
  {year} {2016})}\BibitemShut {NoStop}%
\bibitem [{\citenamefont {Umrigar}\ \emph {et~al.}(2007)\citenamefont
  {Umrigar}, \citenamefont {Toulouse}, \citenamefont {Filippi}, \citenamefont
  {Sorella},\ and\ \citenamefont {Hennig}}]{Umrigar2007}%
  \BibitemOpen
  \bibfield  {author} {\bibinfo {author} {\bibfnamefont {C.~J.}\ \bibnamefont
  {Umrigar}}, \bibinfo {author} {\bibfnamefont {J.}~\bibnamefont {Toulouse}},
  \bibinfo {author} {\bibfnamefont {C.}~\bibnamefont {Filippi}}, \bibinfo
  {author} {\bibfnamefont {S.}~\bibnamefont {Sorella}},\ and\ \bibinfo {author}
  {\bibfnamefont {R.~G.}\ \bibnamefont {Hennig}},\ }\bibfield  {title}
  {\bibinfo {title} {Alleviation of the fermion-sign problem by optimization of
  many-body wave functions},\ }\href@noop {} {\bibfield  {journal} {\bibinfo
  {journal} {Phys. Rev. Lett.}\ }\textbf {\bibinfo {volume} {98}},\ \bibinfo
  {pages} {110201} (\bibinfo {year} {2007})}\BibitemShut {NoStop}%
\bibitem [{\citenamefont {Foulkes}\ \emph {et~al.}(2001)\citenamefont
  {Foulkes}, \citenamefont {Mitas}, \citenamefont {Needs},\ and\ \citenamefont
  {Rajagopal}}]{Matthew2001}%
  \BibitemOpen
  \bibfield  {author} {\bibinfo {author} {\bibfnamefont {W.~M.~C.}\
  \bibnamefont {Foulkes}}, \bibinfo {author} {\bibfnamefont {L.}~\bibnamefont
  {Mitas}}, \bibinfo {author} {\bibfnamefont {R.~J.}\ \bibnamefont {Needs}},\
  and\ \bibinfo {author} {\bibfnamefont {G.}~\bibnamefont {Rajagopal}},\
  }\bibfield  {title} {\bibinfo {title} {Quantum monte carlo simulations of
  solids},\ }\href@noop {} {\bibfield  {journal} {\bibinfo  {journal} {Rev.
  Mod. Phys.}\ }\textbf {\bibinfo {volume} {73}},\ \bibinfo {pages} {33}
  (\bibinfo {year} {2001})}\BibitemShut {NoStop}%
\bibitem [{\citenamefont {Becca}\ and\ \citenamefont
  {Sorella}(2017)}]{BeccaSorella}%
  \BibitemOpen
  \bibfield  {author} {\bibinfo {author} {\bibfnamefont {F.}~\bibnamefont
  {Becca}}\ and\ \bibinfo {author} {\bibfnamefont {S.}~\bibnamefont
  {Sorella}},\ }\href@noop {} {\emph {\bibinfo {title} {Quantum Monte Carlo
  approches for correlated systems}}}\ (\bibinfo  {publisher} {Cambridge
  University},\ \bibinfo {address} {Cambridge, UK},\ \bibinfo {year}
  {2017})\BibitemShut {NoStop}%
\bibitem [{\citenamefont {Anderson}(1976)}]{Anderson1976}%
  \BibitemOpen
  \bibfield  {author} {\bibinfo {author} {\bibfnamefont {J.~B.}\ \bibnamefont
  {Anderson}},\ }\bibfield  {title} {\bibinfo {title} {Quantum chemistry by
  random walk.},\ }\href@noop {} {\bibfield  {journal} {\bibinfo  {journal} {J.
  Chem. Phys.}\ }\textbf {\bibinfo {volume} {65}},\ \bibinfo {pages} {4121}
  (\bibinfo {year} {1976})}\BibitemShut {NoStop}%
\bibitem [{\citenamefont {Marchi}\ \emph {et~al.}(2009)\citenamefont {Marchi},
  \citenamefont {Azadi}, \citenamefont {Casula},\ and\ \citenamefont
  {Sorella}}]{Marchi2009}%
  \BibitemOpen
  \bibfield  {author} {\bibinfo {author} {\bibfnamefont {M.}~\bibnamefont
  {Marchi}}, \bibinfo {author} {\bibfnamefont {S.}~\bibnamefont {Azadi}},
  \bibinfo {author} {\bibfnamefont {M.}~\bibnamefont {Casula}},\ and\ \bibinfo
  {author} {\bibfnamefont {S.}~\bibnamefont {Sorella}},\ }\bibfield  {title}
  {\bibinfo {title} {Resonating valence bond wave function with molecular
  orbitals: Application to first-row molecules},\ }\href@noop {} {\bibfield
  {journal} {\bibinfo  {journal} {J. Chem. Phys.}\ }\textbf {\bibinfo {volume}
  {131}},\ \bibinfo {pages} {154116} (\bibinfo {year} {2009})}\BibitemShut
  {NoStop}%
\bibitem [{\citenamefont {Nakano}\ \emph {et~al.}(2020)\citenamefont {Nakano},
  \citenamefont {Attaccalite}, \citenamefont {Barborini}, \citenamefont
  {Capriotti}, \citenamefont {Casula}, \citenamefont {Coccia}, \citenamefont
  {Dagrada}, \citenamefont {Genovese}, \citenamefont {Luo}, \citenamefont
  {Mazzola}, \citenamefont {Zen},\ and\ \citenamefont {Sorella}}]{TurboRVB}%
  \BibitemOpen
  \bibfield  {author} {\bibinfo {author} {\bibfnamefont {K.}~\bibnamefont
  {Nakano}}, \bibinfo {author} {\bibfnamefont {C.}~\bibnamefont {Attaccalite}},
  \bibinfo {author} {\bibfnamefont {M.}~\bibnamefont {Barborini}}, \bibinfo
  {author} {\bibfnamefont {L.}~\bibnamefont {Capriotti}}, \bibinfo {author}
  {\bibfnamefont {M.}~\bibnamefont {Casula}}, \bibinfo {author} {\bibfnamefont
  {E.}~\bibnamefont {Coccia}}, \bibinfo {author} {\bibfnamefont
  {M.}~\bibnamefont {Dagrada}}, \bibinfo {author} {\bibfnamefont
  {C.}~\bibnamefont {Genovese}}, \bibinfo {author} {\bibfnamefont
  {Y.}~\bibnamefont {Luo}}, \bibinfo {author} {\bibfnamefont {G.}~\bibnamefont
  {Mazzola}}, \bibinfo {author} {\bibfnamefont {A.}~\bibnamefont {Zen}},\ and\
  \bibinfo {author} {\bibfnamefont {S.}~\bibnamefont {Sorella}},\ }\bibfield
  {title} {\bibinfo {title} {\text{TurboRVB}: A many-body toolkit for ab initio
  electronic simulations by quantum monte carlo},\ }\href@noop {} {\bibfield
  {journal} {\bibinfo  {journal} {J. Chem. Phys.}\ }\textbf {\bibinfo {volume}
  {152}},\ \bibinfo {pages} {204121} (\bibinfo {year} {2020})}\BibitemShut
  {NoStop}%
\bibitem [{\citenamefont {Azadi}\ \emph {et~al.}(2015)\citenamefont {Azadi},
  \citenamefont {Singh},\ and\ \citenamefont {K\"uhne}}]{OzoneTurboRVB}%
  \BibitemOpen
  \bibfield  {author} {\bibinfo {author} {\bibfnamefont {S.}~\bibnamefont
  {Azadi}}, \bibinfo {author} {\bibfnamefont {R.}~\bibnamefont {Singh}},\ and\
  \bibinfo {author} {\bibfnamefont {T.~D.}\ \bibnamefont {K\"uhne}},\
  }\bibfield  {title} {\bibinfo {title} {Resonating valence bond quantum monte
  carlo: Application to the ozone molecule},\ }\href@noop {} {\bibfield
  {journal} {\bibinfo  {journal} {Int. J. Quantum Chem.}\ }\textbf {\bibinfo
  {volume} {115}},\ \bibinfo {pages} {1673} (\bibinfo {year}
  {2015})}\BibitemShut {NoStop}%
\bibitem [{\citenamefont {Pritchard}\ \emph {et~al.}(2019)\citenamefont
  {Pritchard}, \citenamefont {Altarawy}, \citenamefont {Didier}, \citenamefont
  {Gibson},\ and\ \citenamefont {Windus}}]{ccp}%
  \BibitemOpen
  \bibfield  {author} {\bibinfo {author} {\bibfnamefont {B.~P.}\ \bibnamefont
  {Pritchard}}, \bibinfo {author} {\bibfnamefont {D.}~\bibnamefont {Altarawy}},
  \bibinfo {author} {\bibfnamefont {B.}~\bibnamefont {Didier}}, \bibinfo
  {author} {\bibfnamefont {T.~D.}\ \bibnamefont {Gibson}},\ and\ \bibinfo
  {author} {\bibfnamefont {T.~L.}\ \bibnamefont {Windus}},\ }\bibfield  {title}
  {\bibinfo {title} {A new basis set exchange: An open, up-to-date resource for
  the molecular sciences community},\ }\href@noop {} {\bibfield  {journal}
  {\bibinfo  {journal} {J. Chem. Inf. Model.}\ }\textbf {\bibinfo {volume}
  {59}},\ \bibinfo {pages} {4814} (\bibinfo {year} {2019})}\BibitemShut
  {NoStop}%
\bibitem [{\citenamefont {Annaberdiyev}\ \emph {et~al.}(2018)\citenamefont
  {Annaberdiyev}, \citenamefont {Wang}, \citenamefont {Melton}, \citenamefont
  {Bennett}, \citenamefont {Shulenburger},\ and\ \citenamefont
  {Mitas}}]{ccECP1}%
  \BibitemOpen
  \bibfield  {author} {\bibinfo {author} {\bibfnamefont {A.}~\bibnamefont
  {Annaberdiyev}}, \bibinfo {author} {\bibfnamefont {G.}~\bibnamefont {Wang}},
  \bibinfo {author} {\bibfnamefont {C.~A.}\ \bibnamefont {Melton}}, \bibinfo
  {author} {\bibfnamefont {M.~C.}\ \bibnamefont {Bennett}}, \bibinfo {author}
  {\bibfnamefont {L.}~\bibnamefont {Shulenburger}},\ and\ \bibinfo {author}
  {\bibfnamefont {L.}~\bibnamefont {Mitas}},\ }\bibfield  {title} {\bibinfo
  {title} {A new generation of effective core potentials from correlated
  calculations: 3d transition metal series},\ }\href@noop {} {\bibfield
  {journal} {\bibinfo  {journal} {J. Chem. Phys.}\ }\textbf {\bibinfo {volume}
  {149}},\ \bibinfo {pages} {134108} (\bibinfo {year} {2018})}\BibitemShut
  {NoStop}%
\bibitem [{\citenamefont {Bennett}\ \emph {et~al.}(2017)\citenamefont
  {Bennett}, \citenamefont {Melton}, \citenamefont {Annaberdiyev},
  \citenamefont {Wang}, \citenamefont {Shulenburger},\ and\ \citenamefont
  {Mitas}}]{ccECP2}%
  \BibitemOpen
  \bibfield  {author} {\bibinfo {author} {\bibfnamefont {M.~C.}\ \bibnamefont
  {Bennett}}, \bibinfo {author} {\bibfnamefont {C.~A.}\ \bibnamefont {Melton}},
  \bibinfo {author} {\bibfnamefont {A.}~\bibnamefont {Annaberdiyev}}, \bibinfo
  {author} {\bibfnamefont {G.}~\bibnamefont {Wang}}, \bibinfo {author}
  {\bibfnamefont {L.}~\bibnamefont {Shulenburger}},\ and\ \bibinfo {author}
  {\bibfnamefont {L.}~\bibnamefont {Mitas}},\ }\bibfield  {title} {\bibinfo
  {title} {A new generation of effective core potentials for correlated
  calculations},\ }\href@noop {} {\bibfield  {journal} {\bibinfo  {journal} {J.
  Chem. Phys.}\ }\textbf {\bibinfo {volume} {147}},\ \bibinfo {pages} {224106}
  (\bibinfo {year} {2017})}\BibitemShut {NoStop}%
\bibitem [{\citenamefont {Azadi}\ \emph {et~al.}(2010)\citenamefont {Azadi},
  \citenamefont {Cavazzoni},\ and\ \citenamefont {Sorella}}]{Azadi2010}%
  \BibitemOpen
  \bibfield  {author} {\bibinfo {author} {\bibfnamefont {S.}~\bibnamefont
  {Azadi}}, \bibinfo {author} {\bibfnamefont {C.}~\bibnamefont {Cavazzoni}},\
  and\ \bibinfo {author} {\bibfnamefont {S.}~\bibnamefont {Sorella}},\
  }\bibfield  {title} {\bibinfo {title} {Systematically convergent method for
  accurate total energy calculations with localized atomic orbitals},\
  }\href@noop {} {\bibfield  {journal} {\bibinfo  {journal} {Phys. Rev. B}\
  }\textbf {\bibinfo {volume} {82}},\ \bibinfo {pages} {125112} (\bibinfo
  {year} {2010})}\BibitemShut {NoStop}%
\bibitem [{\citenamefont {Perdew}\ and\ \citenamefont {Zunger}(1981)}]{lda}%
  \BibitemOpen
  \bibfield  {author} {\bibinfo {author} {\bibfnamefont {J.~P.}\ \bibnamefont
  {Perdew}}\ and\ \bibinfo {author} {\bibfnamefont {A.}~\bibnamefont
  {Zunger}},\ }\bibfield  {title} {\bibinfo {title} {Self-interaction
  correction to density-functional approximations for many-electron systems},\
  }\href@noop {} {\bibfield  {journal} {\bibinfo  {journal} {Phys. Rev. B}\
  }\textbf {\bibinfo {volume} {23}},\ \bibinfo {pages} {5048} (\bibinfo {year}
  {1981})}\BibitemShut {NoStop}%
\bibitem [{\citenamefont {Fahy}\ \emph {et~al.}(1990)\citenamefont {Fahy},
  \citenamefont {Wang},\ and\ \citenamefont {Louie}}]{Fahy90}%
  \BibitemOpen
  \bibfield  {author} {\bibinfo {author} {\bibfnamefont {S.}~\bibnamefont
  {Fahy}}, \bibinfo {author} {\bibfnamefont {X.~W.}\ \bibnamefont {Wang}},\
  and\ \bibinfo {author} {\bibfnamefont {S.~G.}\ \bibnamefont {Louie}},\
  }\bibfield  {title} {\bibinfo {title} {Variational quantum monte carlo
  nonlocal pseudopotential approach to solids: Formulation and application to
  diamond, graphite, and silicon},\ }\href@noop {} {\bibfield  {journal}
  {\bibinfo  {journal} {Phys. Rev. B}\ }\textbf {\bibinfo {volume} {42}},\
  \bibinfo {pages} {3503} (\bibinfo {year} {1990})}\BibitemShut {NoStop}%
\bibitem [{\citenamefont {Ceperley}(1978)}]{Ceperley78}%
  \BibitemOpen
  \bibfield  {author} {\bibinfo {author} {\bibfnamefont {D.~M.}\ \bibnamefont
  {Ceperley}},\ }\bibfield  {title} {\bibinfo {title} {Ground state of the
  fermion one-component plasma: A monte carlo study in two and three
  dimensions},\ }\href@noop {} {\bibfield  {journal} {\bibinfo  {journal}
  {Phys. Rev. B}\ }\textbf {\bibinfo {volume} {18}},\ \bibinfo {pages} {3126}
  (\bibinfo {year} {1978})}\BibitemShut {NoStop}%
\bibitem [{\citenamefont {Casula}(2006)}]{Casula2006}%
  \BibitemOpen
  \bibfield  {author} {\bibinfo {author} {\bibfnamefont {M.}~\bibnamefont
  {Casula}},\ }\bibfield  {title} {\bibinfo {title} {Beyond the locality
  approximation in the standard diffusion monte carlo method},\ }\href@noop {}
  {\bibfield  {journal} {\bibinfo  {journal} {Phys. Rev. B}\ }\textbf {\bibinfo
  {volume} {74}},\ \bibinfo {pages} {161102(R)} (\bibinfo {year}
  {2006})}\BibitemShut {NoStop}%
\bibitem [{\citenamefont {Burke}\ \emph {et~al.}(2016)\citenamefont {Burke},
  \citenamefont {Cancio}, \citenamefont {Gould},\ and\ \citenamefont
  {Pittalis}}]{Burke2016}%
  \BibitemOpen
  \bibfield  {author} {\bibinfo {author} {\bibfnamefont {K.}~\bibnamefont
  {Burke}}, \bibinfo {author} {\bibfnamefont {A.}~\bibnamefont {Cancio}},
  \bibinfo {author} {\bibfnamefont {T.}~\bibnamefont {Gould}},\ and\ \bibinfo
  {author} {\bibfnamefont {S.}~\bibnamefont {Pittalis}},\ }\bibfield  {title}
  {\bibinfo {title} {Locality of correlation in density functional theory},\
  }\href@noop {} {\bibfield  {journal} {\bibinfo  {journal} {J. CHem. Phys.}\
  }\textbf {\bibinfo {volume} {145}},\ \bibinfo {pages} {054112} (\bibinfo
  {year} {2016})}\BibitemShut {NoStop}%
\bibitem [{\citenamefont {McCarthy}\ and\ \citenamefont
  {Thakkar}(2011)}]{McCarthy}%
  \BibitemOpen
  \bibfield  {author} {\bibinfo {author} {\bibfnamefont {S.~P.}\ \bibnamefont
  {McCarthy}}\ and\ \bibinfo {author} {\bibfnamefont {A.~J.}\ \bibnamefont
  {Thakkar}},\ }\bibfield  {title} {\bibinfo {title} {Accurate all-electron
  correlation energies for the closed-shell atoms from \text{Ar} to \text{Rn}
  and their relationship to the corresponding \text{MP2} correlation
  energies},\ }\href@noop {} {\bibfield  {journal} {\bibinfo  {journal} {J.
  Chem. Phys.}\ }\textbf {\bibinfo {volume} {134}},\ \bibinfo {pages} {044102}
  (\bibinfo {year} {2011})}\BibitemShut {NoStop}%
\bibitem [{\citenamefont {Trail}\ and\ \citenamefont {Needs}(2015)}]{CEPP}%
  \BibitemOpen
  \bibfield  {author} {\bibinfo {author} {\bibfnamefont {J.~R.}\ \bibnamefont
  {Trail}}\ and\ \bibinfo {author} {\bibfnamefont {R.~J.}\ \bibnamefont
  {Needs}},\ }\bibfield  {title} {\bibinfo {title} {Correlated electron
  pseudopotentials for 3d-transition metals},\ }\href@noop {} {\bibfield
  {journal} {\bibinfo  {journal} {J. Chem. Phys.}\ }\textbf {\bibinfo {volume}
  {142}},\ \bibinfo {pages} {064110} (\bibinfo {year} {2015})}\BibitemShut
  {NoStop}%
\bibitem [{\citenamefont {Trail}\ and\ \citenamefont {Needs}(2017)}]{eCEPP}%
  \BibitemOpen
  \bibfield  {author} {\bibinfo {author} {\bibfnamefont {J.~R.}\ \bibnamefont
  {Trail}}\ and\ \bibinfo {author} {\bibfnamefont {R.~J.}\ \bibnamefont
  {Needs}},\ }\bibfield  {title} {\bibinfo {title} {Shape and energy consistent
  pseudopotentials for correlated electron systems},\ }\href@noop {} {\bibfield
   {journal} {\bibinfo  {journal} {J. Chem. Phys.}\ }\textbf {\bibinfo {volume}
  {146}},\ \bibinfo {pages} {204107} (\bibinfo {year} {2017})}\BibitemShut
  {NoStop}%
\bibitem [{\citenamefont {Burkatzki}\ \emph {et~al.}(2007)\citenamefont
  {Burkatzki}, \citenamefont {Filippi},\ and\ \citenamefont {Dolg}}]{BurkPP1}%
  \BibitemOpen
  \bibfield  {author} {\bibinfo {author} {\bibfnamefont {M.}~\bibnamefont
  {Burkatzki}}, \bibinfo {author} {\bibfnamefont {C.}~\bibnamefont {Filippi}},\
  and\ \bibinfo {author} {\bibfnamefont {M.}~\bibnamefont {Dolg}},\ }\bibfield
  {title} {\bibinfo {title} {Energy-consistent pseudopotentials for qmc
  calculations},\ }\href@noop {} {\bibfield  {journal} {\bibinfo  {journal} {J.
  Chem. Phys.}\ }\textbf {\bibinfo {volume} {126}},\ \bibinfo {pages} {234105}
  (\bibinfo {year} {2007})}\BibitemShut {NoStop}%
\bibitem [{\citenamefont {Burkatzki}\ \emph {et~al.}(2008)\citenamefont
  {Burkatzki}, \citenamefont {Filippi},\ and\ \citenamefont {Dolg}}]{BurkPP2}%
  \BibitemOpen
  \bibfield  {author} {\bibinfo {author} {\bibfnamefont {M.}~\bibnamefont
  {Burkatzki}}, \bibinfo {author} {\bibfnamefont {C.}~\bibnamefont {Filippi}},\
  and\ \bibinfo {author} {\bibfnamefont {M.}~\bibnamefont {Dolg}},\ }\bibfield
  {title} {\bibinfo {title} {Energy-consistent small-core pseudopotentials for
  3d-transition metals adapted to quantum monte carlo calculations},\
  }\href@noop {} {\bibfield  {journal} {\bibinfo  {journal} {J. Chem. Phys.}\
  }\textbf {\bibinfo {volume} {129}},\ \bibinfo {pages} {164115} (\bibinfo
  {year} {2008})}\BibitemShut {NoStop}%
\bibitem [{Sup()}]{Suppl}%
  \BibitemOpen
  \href@noop {} {}\bibinfo {note} {"See Supplemental Material."}\BibitemShut
  {NoStop}%
\bibitem [{\citenamefont {Azadi}\ and\ \citenamefont
  {Foulkes}(2015)}]{Azadi2015}%
  \BibitemOpen
  \bibfield  {author} {\bibinfo {author} {\bibfnamefont {S.}~\bibnamefont
  {Azadi}}\ and\ \bibinfo {author} {\bibfnamefont {W.}~\bibnamefont
  {Foulkes}},\ }\bibfield  {title} {\bibinfo {title} {Systematic study of
  finite-size effects in quantum monte carlo calculations of real metallic
  systems},\ }\href@noop {} {\bibfield  {journal} {\bibinfo  {journal} {J.
  Chem. Phys.}\ }\textbf {\bibinfo {volume} {143}},\ \bibinfo {pages} {102807}
  (\bibinfo {year} {2015})}\BibitemShut {NoStop}%
\end{thebibliography}%

\end{document}